%% file: dune.tex
\documentclass[11pt,a4paper]{article}
\pdfoutput=1
\usepackage[utf8]{inputenc}
\usepackage[T1]{fontenc}
\usepackage{jcappub}
\usepackage{bm}
\usepackage{booktabs}
\usepackage{array}
\usepackage[section]{placeins}

\newcommand{\nue}{\nu_e}
\newcommand{\dmsq}{\Delta m^2_{31}}
\newcommand{\Losc}{L_{\rm osc}^{\rm res}}
\newcommand{\Pee}{P_{ee}}
\newcommand{\sthirteen}{\sin^2\theta_{13}}

\title{\bf Numerical shock broadening and supernova neutrino conversion:
implications for DUNE}

\author[a]{Rafael C. R. de Lima,}
\author[a]{Pedro Martins Bailer}

\affiliation[a]{Departamento de F\'isica, Centro de Ci\^encias Tecnol\'ogicas,\\
Universidade do Estado de Santa Catarina (UDESC),\\
Rua Paulo Malschitzki 200, 89219-710 Joinville, SC, Brazil\\
ORCID (R.C.R.L.): \href{https://orcid.org/0000-0003-1718-3838}{0000-0003-1718-3838}}

\emailAdd{rafael.nuclear@gmail.com}
\emailAdd{pedromartinsbailer@gmail.com}


\abstract{Hydrodynamical simulations spread the supernova shock front over a few grid zones,
a width set by the numerical scheme rather than by the physics of the shock. Where the front
crosses the H resonance, that width can make the computed MSW evolution artificially
adiabatic. We quantify the effect on six three-dimensional blue-supergiant explosion models
from the Garching archive by propagating individual rays twice: through the stored profile,
and with the front replaced by a narrow reconstruction, $2.2$~km wide, that approximates a
sudden crossing while keeping the densities on either side. The stored fronts are $1.6$ to
$1.7$ grid zones wide. Across twelve configurations, each chosen inside the energy band in
which that model's front meets the resonance, the reconstruction raises the ray-averaged
crossing probability $P_H$ by $0.44$ to $0.76$ in absolute terms, median $0.56$; narrowing the
reconstruction thirty-fold changes each difference by at most $0.055$. Along all
$12\,878$ rays of one configuration, the sudden-limit formula evaluated on each ray's own
densities agrees closely with the propagated reconstructed result. The change is largest when the
density jump spans the resonance, over a band of energies that can be estimated from the
presupernova density once the shock radius and the compression are known, with tails outside
it. In a separate calculation with a parametrised shock and a published
emission model, time-dependent conversion changes the $\nue$ charged-current count in a
$40$~kt argon detector by $-4.4\%$, with a sign set by the flavour composition of the source.
Folding the reconstructed three-dimensional fronts with a consistent emission model remains
to be done.}

\begin{document}
\maketitle
\flushbottom

\section{Introduction}
\label{sec:intro}

A galactic core-collapse supernova will deposit of order $10^3$ neutrino events in DUNE
\citep{dune2021}, and unlike the water-Cherenkov, scintillator and ice detectors, which are
dominated by $\bar\nu_e$, liquid argon responds mainly to $\nue$ through
$\nue + {}^{40}\mathrm{Ar} \to e^- + {}^{40}\mathrm{K}^*$. That is the channel in which the H
resonance acts in normal ordering: adiabatic conversion gives $\Pee = \sthirteen = 0.022$
there, against $\sin^2\theta_{12}\cos^2\theta_{13} = 0.30$ in inverted ordering
\citep{dighe2000}, and the difference reaches the count to the extent that the $\nue$ and
$\nu_x$ fluxes differ. For the first seconds the shock is crossing the envelope in which the
H resonance sits, so the conversion depends on time and on the structure of the front
\citep{schirato2002,fogli2003,tomas2004}.

However, hydrodynamical simulations do not resolve the physical shock structure. A shock in a
simulation
is spread over a few grid zones, a width set by the numerical scheme, and at the resonance
radius a few zones can be several times the oscillation length $\Losc$ that governs the
crossing. A front that wide is crossed adiabatically. Propagating neutrinos through a
simulated profile as stored can therefore suppress the conversion that a thin physical shock
would produce.

Neither the effect nor its remedy is new. Shock-induced non-adiabaticity has been studied
since Schirato \& Fuller~\cite{schirato2002}, with the phenomenology developed by Fogli
\emph{et al.}~\cite{fogli2003} and Tom\`as \emph{et al.}~\cite{tomas2004}, the mechanism
formalised as stimulated transitions by Kneller, McLaughlin \&
Patton~\cite{kneller2013sinusoidal} and Patton, Kneller \&
McLaughlin~\cite{patton2014,patton2015}, and the consequences for a liquid argon detector
examined by Gil-Botella \& Rubbia~\cite{gilbotella2003}. When profiles were one-dimensional
and handled by hand, the numerical front was steepened before the flavour evolution was
computed \citep{tomas2004,barger2005}. Detector predictions from three-dimensional models now
commonly apply a constant instead --- {\sc SNEWPY} \citep{snewpy2021}, which drives
{\sc SNOwGLoBES} \citep{snowglobes}, offers for standard MSW in its release~1.5.1 only
$P_H = 0$ and $P_H = 1$, and Choi, Burrows \& Vartanyan~\cite{choi2025}, for example, pass $24$
three-dimensional models through the adiabatic prescription. To our knowledge, the impact of numerical shock broadening on neutrino flavour conversion has
not been quantified systematically using three-dimensional supernova models.

This work makes that measurement as a controlled numerical comparison. Along individual rays
of six three-dimensional blue-supergiant models we propagate the same profile twice, once as
stored and once with the front replaced by a narrow reconstruction that approximates a sudden
crossing, and take the difference in the crossing probability $P_H$ (Section~\ref{sec:3d}).
Across twelve configurations the difference is $0.44$ to $0.76$ in absolute terms, and along
all rays of one configuration a closed form agrees closely with the propagated result.
Section~\ref{sec:quando} establishes where the comparison matters --- most where the density
jump spans the resonance, over a band of energies that can be estimated without a flavour
calculation --- and what limits it: the width of the reconstruction, the epoch, and the
uncertainty in placing the reconstructed front.
Section~\ref{sec:estruturas} summarises perturbative tests of smoother structures and of
turbulence, whose conclusions are conditional on the assumptions stated there.

Whether a change in $P_H$ moves the event count, and in which direction, depends on the
flavour composition of the flux, and folding the three-dimensional fronts with a consistent
emission model is beyond this paper. Section~\ref{sec:dune} gives instead a separate
demonstration with a parametrised front and a published emission model.
Section~\ref{sec:metodo} states the method and Section~\ref{sec:limites} the limitations.

\section{Method}
\label{sec:metodo}

\subsection{Evolution and observable}
\label{sec:evolucao}

We integrate three flavours along radial trajectories with the matter Hamiltonian
\citep{wolfenstein1978,ms1985}
\begin{equation}
  H(r) \;=\; \frac{1}{2E}\,U\,
  \mathrm{diag}\!\left(0,\ \Delta m^2_{21},\ \dmsq\right)U^{\dagger}
  \;+\; \mathrm{diag}\!\left(\sqrt{2}\,G_F\,Y_e\,\rho(r)/m_n,\,0,\,0\right),
  \label{eq:H}
\end{equation}
with mixing parameters close to those of Esteban \emph{et al.}~\cite{nufit2020}, whose values
are listed with the rest of the protocol in Table~\ref{tab:protocolo}; the choice of the
neutron mass in the matter term is discussed in Appendix~\ref{ap:validacao}. The Hamiltonian
is assembled and propagated by {\sc GHOST} \citep{ghost}, an open-source code written for this
work.\footnote{\url{https://github.com/RafaelCRdeLima/GHOST}, under the MIT licence.} The evolution
operator over each radial slice is expanded in Chebyshev polynomials of the Hamiltonian
\citep{tallallqvist2007} with rigorous spectral bounds, and slices are bisected adaptively: a
slice is accepted when the state vectors obtained with $k$ and $2k$ substeps agree, component
by component, in the complex amplitude rather than in the populations, and the finer of the
two solutions is kept. The solver agrees with the independent exact code {\sc NuOscProbExact}
\citep{bustamante2019} to $2.2\times10^{-14}$ over $320$ vacuum and constant-density
configurations. This validates the implementation of the Hamiltonian and mixing parameters; it is not by
itself a statement
about a varying profile, for which Appendix~\ref{ap:validacao} gives a closed-form comparison
on a profile with a gradient. The solver is exposed as a {\sc SNEWPY}
\texttt{FlavorTransformation} (\texttt{tools/neutrinos/snewpy\_ghost.py} in the {\sc GHOST}
repository), so that the rest of the pipeline is unchanged.

The observable is the incoherent combination $\Pee = \sum_i |U_{ei}|^2 P_i$. The evolution
itself is coherent by assumption: amplitudes are propagated through the whole profile, and
every phase accumulated between features, including between successive crossings of the
front, is present in the $P_i$. The incoherent projection is taken once, at the end, and stands
for the separation of the mass eigenstate wave packets on the way to the detector
\citep{kersten2016}. With $\sigma_x \simeq 10^{-11}$~cm at production the packets separate by
one width in $12$ to $35$~km. The features between which the calculation retains interference
are much further apart: the two edges of the compressed shell of the parametrised front are
$0.15\,r_f$ apart, $3000$~km at $t = 1$~s, and crossings in the three-dimensional profiles are
up to $3\times10^4$~km apart. The coherent treatment therefore retains interference that
decoherence could reduce; Kersten \& Smirnov~\cite{kersten2016} also show that a sharp density
change can partially restore coherence. Settling it would need the separation of the packets
computed along the relevant trajectory, or the density matrix propagated with the packet
structure retained, which we have not done. Everything below is a coherent calculation. The main
result does not rest on it where it was tested ray by ray (Section~\ref{sec:subita}): there the
reconstructed value is the single-step sudden result, which involves no interference between
separated features. What depends on the assumption are secondary structures --- the modulation
below $0.1\,\Losc$ in Table~\ref{tab:largura} and the local maxima within the band of
Figure~\ref{fig:mapa} --- and, in the early-epoch configurations, where most rays cross the
resonance more than once, any interference between those crossings.

\subsection{Crossing probability}
\label{sec:ph}

We quote the crossing probability $P_H$ by two routes, and each table says which.
Table~\ref{tab:tresvias} uses the mass-basis population, $P_H = 1 - m_3$;
Table~\ref{tab:orcamento} inverts the incoherent observable,
$P_H = (\Pee - D_3)/(D_2 - D_3)$ with $D_i = |U_{ei}|^2$. With the final populations
normalised the two are related by an identity,
\begin{equation}
  P_H^{\rm inc} - P_H^{\rm mass} \;=\; \frac{D_1 - D_2}{D_2 - D_3}\, m_1
  \;=\; 1.4061\, m_1 ,
  \label{eq:duasdef}
\end{equation}
so they agree to the extent that nothing reaches the light state. Over the $20$ rays of each of
the twelve configurations of Table~\ref{tab:tresvias}, raw and reconstructed at every width
tested, the gap is at most $9.4\times10^{-3}$, at most $2\%$ of the difference reported for that
configuration. The tables state which definition they use, and are not interchanged.

\subsection{Oscillation length and resonance}
\label{sec:losc}

The scale against which a structure at the resonance is measured is the oscillation length
there, where the splitting between matter eigenstates is minimal,
\begin{equation}
  \Losc \;=\; \frac{4\pi E}{\dmsq \sin 2\theta_{13}} \;=\; 66.8\ \mathrm{km}
  \qquad (E = 20\ \mathrm{MeV}),
  \label{eq:losc}
\end{equation}
which scales as $E$. A structure much narrower than $\Losc$ is crossed suddenly, one much
wider adiabatically. The resonance condition is $Y_e\rho = 792$~g\,cm$^{-3}$ at $20$~MeV. In
the progenitors used here the crossing falls in helium-rich material --- over the six models
at both epochs we measure $Y_e = 0.50$ to $0.52$ and a hydrogen mass fraction below $0.06$ at
the crossing --- so $\rho_{\rm res} \simeq 1590$~g\,cm$^{-3}$; the envelope value
$Y_e \simeq 0.83$ would place it at $954$~g\,cm$^{-3}$.

\subsection{Initial conditions and domain}
\label{sec:dominio}

The domain must start well above the H resonance for the lowest energy and end below the L
resonance for the highest. The three-dimensional files begin at density ratios
$\rho(r_0)/\rho_H$ between $113$ and $1.2$ for the configurations used, so each ray is
extended inward by a power law $\rho \propto r^{-\alpha}$ to a ratio of $10^2$ before
propagation; on these profiles the result flattens once the ratio exceeds $10$. The effect of the extension is quantified in Section~\ref{sec:consequencia}, and the requirements
on the domain are set out in Appendix~\ref{ap:validacao}.

\subsection{Reconstruction of the front}
\label{sec:reconstrucao}

In hydrodynamical simulations the shock is spread over a few grid zones by the numerical
scheme, while the Rankine--Hugoniot jump conditions relate the upstream and downstream states;
no multi-dimensional simulation resolves the physical width.\footnote{We are grateful to H.-Th. Janka for
restating this point to us, and for the observation --- which we had not anticipated --- that
angle-averaged profiles of a deformed three-dimensional shock are smeared further still.}

The physical shock is much thinner. Where the radiation behind a shock dominates the pressure
and the material ahead is opaque enough to trap it, the shock is radiation-mediated, with a
transition layer of optical depth of order $c/v_s$ \citep{levinson2020}. Both conditions hold
on the measured front: the momentum jump at the measured compression and the front speed
inferred from successive snapshots, $v_s = 1.15\times10^{4}$~km\,s$^{-1}$, puts the
downstream temperature near $6\times10^{8}$~K, where radiation pressure exceeds gas pressure
by a factor of about $5$, and with $\kappa\rho \simeq 60$~cm$^{-1}$ ahead of the front a single
grid zone is $10^{9}$ optical depths thick, against the $c/v_s \simeq 26$ needed. The width is
then of order $c/(\kappa\rho v_s)$, with $v_s$ the speed relative to the unshocked material,
which is essentially at rest, and $\kappa = 0.40\,Y_e$~cm$^2$\,g$^{-1} \simeq
0.20$~cm$^2$\,g$^{-1}$ for fully ionised matter at $Y_e \simeq 0.5$. For $\rho$ between
$3\times10^{2}$~g\,cm$^{-3}$ ahead of the front and $1.6\times10^{3}$~g\,cm$^{-3}$ at the
resonance that is $4$ to $0.8$~mm, seven to eight orders of magnitude below $\Losc$. These
are estimates from the jump conditions, not readings of the temperatures stored in the
profiles.

On the scale of $\Losc$ the physical front is therefore effectively a step. We replace the
numerically broadened front with a narrow reconstruction intended to approximate a sudden
crossing, as was the practice of the one-dimensional era, where the front was ``steepened
artificially to reintroduce the physical requirement of a density discontinuity which is lost
in hydrodynamic simulations'' \citep{barger2005}. The procedure, with every parameter, is in
Table~\ref{tab:protocolo}. The zones to replace are the one with the largest drop in
$\ln\rho$ between neighbours and the contiguous zones whose drop exceeds a quarter of it. They
are replaced by a transition $\rho = \rho_{\rm in}(1-s) + \rho_{\rm out}\,s$ with
$s = \tfrac12[1+\tanh((r-r_f)/w)]$, where $\rho_{\rm in}$ and $\rho_{\rm out}$ continue the
stored profile on either side and $w = 1$~km, so that the transition is $2.2$~km wide from
$10$ to $90\%$ --- $0.033\,\Losc$ at $20$~MeV and $0.13\,\Losc$ at $5$~MeV. It is not an exact
step, and Section~\ref{sec:consequencia} quantifies the effect of the remaining width. Hereafter we refer to the profile as stored in the archive file, propagated as it is, as
\emph{raw}, and to the profile with the narrow transition as \emph{reconstructed}. In the figure legends, ``as stored'' and ``raw hydrodynamic profile''
denote the former, and ``sudden-limit reconstruction'' and ``sharpened'' the latter. The density
criterion places the reconstructed front about one grid zone from where the velocity
discontinuity puts it; Section~\ref{sec:visada} quantifies the effect of that offset.

Each propagation uses the profile of one snapshot, frozen while the neutrino crosses it. For a
front moving at $v_f$ this neglects its displacement during the crossing, an approximation of
order $v_f/c$, which is $0.03$ to $0.13$ over the velocities of Section~\ref{sec:mapa}; the
times we quote are those of the snapshot.

\begin{table}[h]
\centering
\small
\begin{tabular}{@{}>{\raggedright\arraybackslash}p{0.30\textwidth}>{\raggedright\arraybackslash}p{0.64\textwidth}@{}}
\toprule
item & value \\
\midrule
initial state & pure $\nue$ ($\bar\nu_e$) at $r_0$; $\Pee$ projected incoherently at the end \\
mixing, three-dimensional rays & $\theta_{12} = 33.4^\circ$, $\theta_{13} = 8.6^\circ$, $\theta_{23} = 49^\circ$, $\delta_{\rm CP} = -90^\circ$, $\Delta m^2_{21} = 7.42\times10^{-5}$~eV$^2$, $\dmsq = 2.51\times10^{-3}$~eV$^2$ \\
mixing, parametrised profile & {\sc SNEWPY} defaults, $\sthirteen = 0.0222$ \\
integration tolerance & $10^{-6}$ on each complex amplitude, $k$ against $2k$ substeps \\
rays per configuration & $20$, drawn uniformly over the cells with seed $7$ \\
inner domain (3D) & extended by $\rho \propto r^{-3}$ to $\rho(r_0)/\rho_H = 10^2$ \\
front selection & largest drop in $\ln\rho$ between neighbouring zones, plus contiguous zones with a drop above $1/4$ of it \\
reconstruction & $\tanh$ transition, $w = 1$~km ($2.2$~km from $10$ to $90\%$; $0.3$, $0.1$ and $0.03$~km as a check), $400$ points over $r_f \pm 12w$; each side log-interpolated from the stored zones outside the transition; $Y_e$ interpolated from the stored profile \\
parametrised profile & $\rho = 10^{10}\,(100\,\mathrm{km}/r)^{3}$~g\,cm$^{-3}$, $Y_e = 0.5$, $600$~km to $5\times10^{5}$~km \\
parametrised front & $r_f = v_f t$, $v_f = 2\times10^{4}$~km\,s$^{-1}$, compression $10$, rarefaction $0.15$, shell $0.15\,r_f$ thick; $1500$ points over $0.7$--$1.3\,r_f$ ($\simeq 8$~km at $t = 1$~s); a step between grid points unless a width is stated \\
\bottomrule
\end{tabular}
\caption{The protocol of the propagations.}
\label{tab:protocolo}
\end{table}

\section{Numerical shock broadening and its effect in six models}
\label{sec:3d}

\subsection{The models and the measured front}
\label{sec:modelos}

\begin{figure}[t]
\centering
\includegraphics[width=\textwidth]{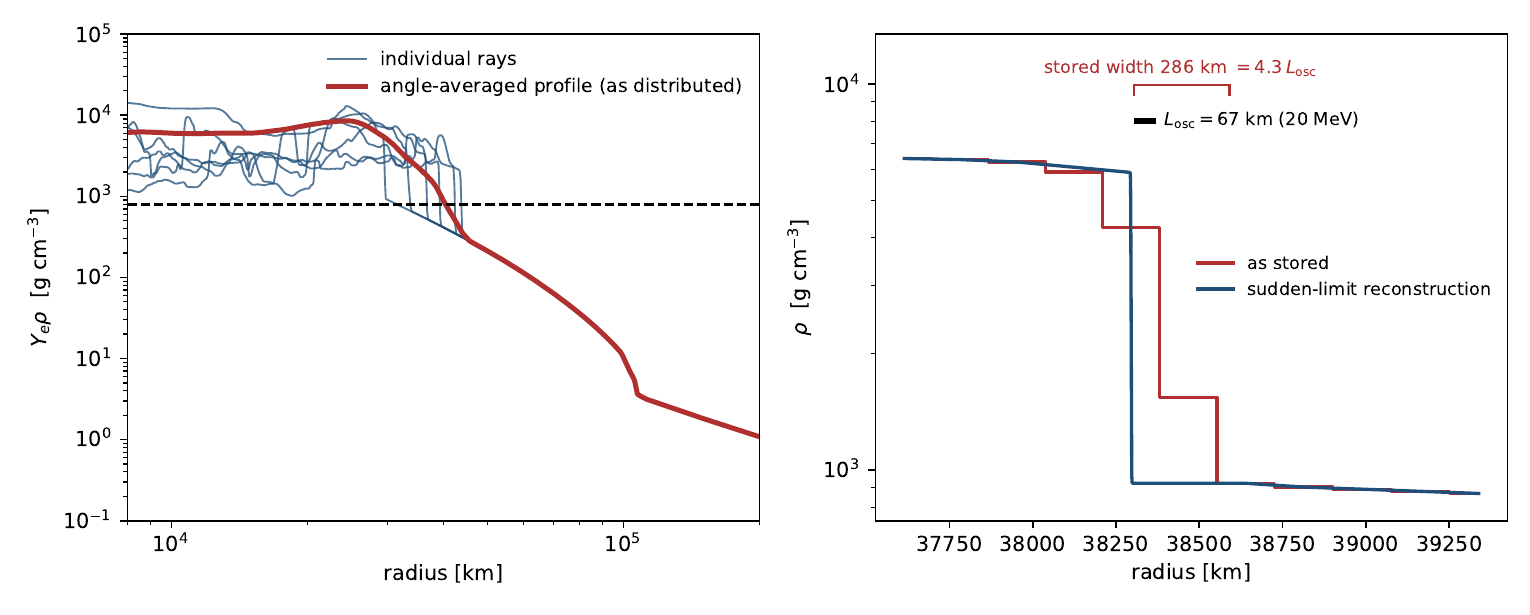}
\caption{\emph{Left:} $Y_e\rho$ along six individual rays of M15-7b at $t = 3$~s, against the
angle-averaged profile as distributed by the archive --- not one we computed --- with the H
resonance condition, $Y_e\rho = 792$~g\,cm$^{-3}$, marked dashed. The shock radius varies by
$38\%$ between directions, so the average smears a front that each ray crosses sharply.
\emph{Right:} density across the front along one ray, as stored and with the narrow
reconstruction of Table~\ref{tab:protocolo}, which at this scale looks like a step. The
bracket marks the stored width, $286$~km, the radial interval over which the velocity falls
from $90\%$ to $10\%$ of its post-shock value; the bar below it, with the same origin, is one
oscillation length at $20$~MeV. The ray shown is the one whose stored width, $4.3\,\Losc$, is
the median over $600$ rays ($10$--$90\%$: $3.2$ to $5.2$). The reconstructed front is placed by
the density criterion, here at the inner end of the bracket, about one zone from the velocity
front; in the one configuration tested no statistically significant shift of the mean was found
(Section~\ref{sec:visada}), but the offset is not negligible along a single ray.}
\label{fig:perfis}
\end{figure}

We use six blue-supergiant explosion models from the Garching Core-Collapse Supernova
Archive:\footnote{\url{https://wwwmpa.mpa-garching.mpg.de/ccsnarchive/}; models discussed in
Wongwathanarat, M\"uller \& Janka~\cite{wongwathanarat2015} and Utrobin
\emph{et al.}~\cite{utrobin2015,utrobin2019,utrobin2021}.} M15-7b of A.~Wongwathanarat, and
W16-3, W18-3, W18r-2, W18x-2 and W20-5 \citep{utrobin2019}. Each is distributed as
angle-averaged radial profiles and as the three-dimensional output on the Yin-Yang grid: two
patches $\times\,137$ in $\varphi$ $\times\,47$ in $\theta$ $\times\,2400$ radial zones, giving
$12\,878$ rays with density, $Y_e$, temperature, velocity and composition. Every result below
uses the three-dimensional data. Per-ray statistics are unweighted averages over grid cells
(Section~\ref{sec:limites}). At $t = 3.0$~s we find, over $400$ sampled rays of M15-7b:

\begin{table}[h]
\centering
\begin{tabular}{lr}
\toprule
shock front radius & $3.478\times10^{4}$~km \\
ray-to-ray spread of the front & $38\%$ of the radius \\
H resonance ($Y_e\rho = 792$) & $3.462\times10^{4}$~km, found in $400/400$ rays \\
compression across the front & $5.22\times$ over the density bracket \\
numerical width of the front & $281$~km $= 4.2\,\Losc(20~\mathrm{MeV})$ \ ($1.7$ zones) \\
\bottomrule
\end{tabular}
\caption{The shock at $t = 3$~s in M15-7b, measured along individual rays. The radius, its
spread and the width locate the front by the velocity discontinuity; the compression is
taken over the bracket of the density detector, and over the narrower velocity bracket the
same rays give $3.7\times$.}
\label{tab:m157b}
\end{table}

At this epoch the front and the H resonance coincide, at $3.478$ against
$3.462\times10^{4}$~km. Measured the same way on all six progenitors --- at $3$~s for M15-7b
and at $1.68$--$1.69$~s for the W models --- the numerical width is $281$, $255$, $235$, $251$,
$312$ and $282$~km for M15-7b, W16-3, W18-3, W18r-2, W18x-2 and W20-5. In grid zones the six
vary by only about $2.5\%$, from $1.63$ to $1.67$, which indicates that the width is set by
the scheme and not by the star. Expressed in the oscillation length at a reference energy of
$20$~MeV these are $3.5$ to $4.7\,\Losc$; since $\Losc \propto E$, at the energies of the
early-epoch configurations of Table~\ref{tab:tresvias} the same widths are $4.2$ to
$14.1\,\Losc$. Locating the front by its largest density drop instead of the velocity
discontinuity widens it, to $5.7\,\Losc(20~\mathrm{MeV})$ in M15-7b. The inferred compression ratio depends on the interval over which it is measured: in M15-7b, $5.22$ over the density
bracket, $3.7$ over the velocity bracket, and $13.9$ over a fixed geometric one, where the
background gradient dominates.

The angle-averaged profiles are wider still. Measured the same way --- the radial interval
over which the velocity falls from $90\%$ to $10\%$ of its post-shock value --- the front of
M15-7b is $2.5\times10^{3}$~km at $0.3$~s, $9.7\times10^{3}$ at $1$~s, $4.5\times10^{4}$ at
$3$~s and $2.0\times10^{4}$ at $10$~s: $37$, $145$, $673$ and $300\,\Losc(20~\mathrm{MeV})$.
Defining the contrast of the front as the peak of $|d\ln\rho/d\ln r|$ divided by its median in
a window of $\pm 0.25$ decades about the front, M15-7b at $t = 3.001$~s gives $50.9$ along an
individual line of sight, measured over all $12\,878$ rays; our own angular average of the
same data gives $2.89$, and the archive's angle-averaged profile $2.70$. The agreement of the
last two shows that the loss is due to the angular averaging. Its cause is geometric: the
front radius disperses by $13.6\%$ as a standard deviation between directions, and averaging
a step whose position varies by $15\,000$~km cannot produce a step. Angle-averaged radial
profiles, which is what most model sets distribute, do not retain the quantity this
comparison depends on.

\subsection{The paired comparison}
\label{sec:consequencia}

Propagating rays with the raw and with the reconstructed profile gives
Table~\ref{tab:tresvias}, with $P_H = 1 - m_3$ (Section~\ref{sec:ph}).

\begin{table}[h]
\centering
\small
\setlength{\tabcolsep}{4pt}
\begin{tabular}{llrrrrlrr}
\toprule
model & $t$ (s) & $E$ (MeV) & $\rho_0/\rho_H$ & $P_H$ raw & $P_H$ rec. & difference
  & $\alpha$ spread & $w = 0.03$~km \\
\midrule
M15-7b & $3.00$ & $20$ & $113$ & $0.002$ & $0.478$ & $\bm{0.476} \pm 0.076$ & $0.002$ & $0.486$ \\
M15-7b & $3.00$ & $10$ & $56$  & $0.002$ & $0.767$ & $\bm{0.765} \pm 0.041$ & $0.005$ & $0.786$ \\
W16-3  & $1.69$ & $10$ & $17$  & $0.215$ & $0.669$ & $\bm{0.454} \pm 0.054$ & $0.045$ & $0.413$ \\
W16-3  & $8.70$ & $60$ & $8$   & $0.065$ & $0.526$ & $\bm{0.461} \pm 0.046$ & $0.018$ & $0.441$ \\
W18-3  & $1.68$ & $5$  & $5$   & $0.062$ & $0.625$ & $\bm{0.564} \pm 0.037$ & $0.011$ & $0.619$ \\
W18-3  & $8.72$ & $60$ & $78$  & $0.025$ & $0.761$ & $\bm{0.736} \pm 0.028$ & $0.005$ & $0.745$ \\
W18r-2 & $1.69$ & $8$  & $11$  & $0.012$ & $0.663$ & $\bm{0.651} \pm 0.028$ & $0.002$ & $0.677$ \\
W18r-2 & $8.67$ & $60$ & $1.2$ & $0.032$ & $0.689$ & $\bm{0.658} \pm 0.025$ & $0.018$ & $0.657$ \\
W18x-2 & $1.68$ & $14$ & $20$  & $0.020$ & $0.627$ & $\bm{0.606} \pm 0.030$ & $0.040$ & $0.614$ \\
W18x-2 & $8.71$ & $100^{\dagger}$ & $1.2$ & $0.074$ & $0.518$ & $\bm{0.444} \pm 0.043$ & $0.045$ & $0.444$ \\
W20-5  & $1.69$ & $10$ & $10$  & $0.186$ & $0.742$ & $\bm{0.556} \pm 0.061$ & $0.056$ & $0.594$ \\
W20-5  & $8.67$ & $60$ & $21$  & $0.074$ & $0.589$ & $\bm{0.515} \pm 0.071$ & $0.052$ & $0.514$ \\
\bottomrule
\end{tabular}
\caption{The same data propagated two ways (protocol in Table~\ref{tab:protocolo}). Each
entry is the \emph{mean} of $P_H = 1 - m_3$ over $20$ rays, with the reconstruction at
$w = 1$~km; the difference carries the standard error of the mean of the paired differences.
$\rho_0/\rho_H$ is the ratio the archive file itself reaches before extension; ``$\alpha$
spread'' is the largest departure of the difference from its $\alpha = 3$ value when $\alpha$
is moved to $2$ or $4$; the last column is the difference with the reconstruction narrowed to
$w = 0.03$~km. Each energy lies
inside that model's band at that epoch (Section~\ref{sec:banda}). $^{\dagger}$W18x-2 at
$8.71$~s has its band at $87$--$457$~MeV, above the emitted spectrum.}
\label{tab:tresvias}
\end{table}

The extension of the domain is a modelling choice, so we quantified its impact. The full
ladder of target ratios --- $10$, $30$, $10^2$, $3\times10^2$, $10^3$ and $3\times10^3$, each
at $\alpha = 2$, $3$ and $4$ --- was run on the two configurations that bracket the problem:
M15-7b at $3$~s and $20$~MeV, whose files offer a ratio of $113$, and W18r-2 at $8.7$~s and
$60$~MeV, whose files offer $1.2$. In the first the difference stays within $0.474$ to $0.480$
over the whole ladder. In the second it moves once, from
$0.309$ with no extension to $0.643$--$0.662$ at a ratio of $10$, and then stays within
$0.629$--$0.666$ out to $3\times10^3$. The other ten configurations were run at a ratio of
$10^2$ at all three $\alpha$, with a $10^3$ check at $\alpha = 3$ that agrees. The spread over
$\alpha$ has a median of $0.018$ and reaches $0.056$, a $4\%$ systematic in the median entry
and $10\%$ in the worst, and the table gives it per entry. In the two entries whose files would
begin on the resonance, W18r-2 and W18x-2 at the late epoch, deepening the domain lowers the
raw column from $0.27$ to $0.03$ and from $0.22$ to $0.07$; in the other ten the raw column
moves by at most $0.005$, and the reconstructed one by up to $0.07$ in either direction,
largest in W16-3 and W20-5 at $1.69$~s and W18x-2 at $1.68$~s.

The difference is $0.44$ to $0.76$, with a median over the twelve configurations of $0.56$,
on a quantity bounded between $0$ and $1$. We report the difference and not the ratio of the
two columns, because the ratio depends on how close the raw value is to zero: in M15-7b the
raw profile returns $P_H = 0.002$ at both energies and the ratio is in the hundreds, while in
W16-3 at $1.69$~s it returns $0.215$ and the ratio is $3.1$. The raw column is a median
$7.3\%$ of the reconstructed one, and at most $32\%$. These are twelve selected
configurations --- energies inside each model's band, two epochs --- and not a statistic over
all energies and epochs or a prediction for a population of events.

The same twenty rays per configuration were propagated again, with every per-ray result
recorded, at the reconstruction width of Table~\ref{tab:protocolo} and at three narrower ones.
At $w = 1$~km the rerun reproduces every mean of Table~\ref{tab:tresvias}. The standard error
of the mean difference is $0.025$ to $0.076$. Narrowing the reconstruction to $w = 0.3$, $0.1$
and $0.03$~km changes the differences by $-0.042$ to $+0.055$, and between $0.1$ and $0.03$~km by
no more than $0.005$ in any configuration, so the narrow limit is reached by $w = 0.1$~km. The
range becomes $0.41$ to $0.79$, with a median of $0.60$. The largest change, $+0.055$, is at
$5$~MeV in W18-3, where $w = 1$~km is the largest fraction of $\Losc$, and it is the only one
that exceeds the standard error of its mean. Three numbers therefore accompany each difference
and measure different things: the standard error, from the dispersion between rays; the
$\alpha$ spread, the sensitivity to the inward extension of the domain; and the change with
$w$, the sensitivity to the residual width of the reconstruction.

Both columns vary between configurations, and we have not isolated the cause of the
variation in the raw one. The rarefaction behind the front lets $\rho_{\rm res}$ be met more
than once: over all $12\,878$ rays, the fraction crossing the resonance exactly once falls from
$97.8\%$ in M15-7b at $3$~s to $39.7\%$, $22.2\%$ and $6.7\%$ in W20-5, W16-3 and W18-3 at
$1.69$~s, and to zero in W18r-2 and W18x-2, so single-crossing statements --- the sudden
formula below and the band of Section~\ref{sec:banda} --- describe a majority of the rays only
at the later epoch. The early- and late-epoch rows do not differ significantly (medians $0.56$
and $0.52$), and since they were run at different energies epoch and energy cannot be
separated here.

\begin{figure}[t]
\centering
\includegraphics[width=\textwidth]{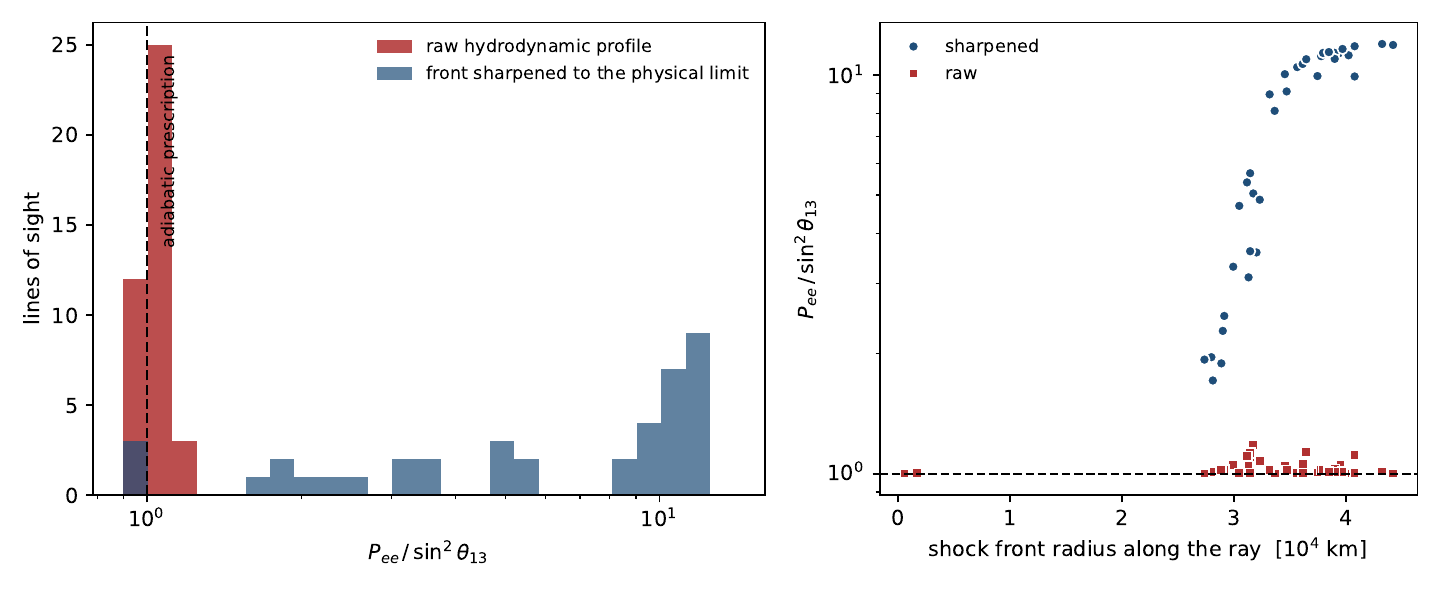}
\caption{The same simulation data propagated two ways, along $40$ lines of sight of M15-7b
at $t = 3$~s and $E = 20$~MeV. \emph{Left:} distribution of $\Pee$ in units of the adiabatic
prescription, raw and reconstructed (``sharpened'' in the legend). \emph{Right:} the same,
against the shock front radius along each ray. Three of the $40$ rays show no effect because
the front detector locates an inner feature rather than the shock; they are kept, and they
lower the median $P_H$ by $0.073$. The spread contains the uncertainty of the reconstruction
as well as physical dispersion (Section~\ref{sec:visada}).}
\label{fig:raios}
\end{figure}

Figure~\ref{fig:raios} shows one configuration ray by ray: the raw values cluster at the
adiabatic prescription, and the reconstructed ones are raised by about $0.5$ in $P_H$ and
spread widely.

\subsection{The sudden limit, ray by ray}
\label{sec:subita}

For a step the crossing probability is given by the sudden approximation,
$P_H = \sin^2(\theta_m^{-} - \theta_m^{+})$, with $\theta_m^{\mp}$ the matter mixing angles on
the two sides of the front, which Fogli \emph{et al.}~\cite{fogli2003} wrote down for the shock.
To test it against propagation along individual lines of sight, we propagated every ray of
M15-7b at $t = 3.00$~s and $E = 20$~MeV --- all $12\,878$, none discarded --- twice each, and
evaluated the formula on each ray from that ray's own densities on the two sides of the
reconstructed front that was actually propagated. No free parameters are fitted in this
comparison.

\begin{table}[h]
\centering
\begin{tabular}{lrrr}
\toprule
 & mean & median & s.d. \\
\midrule
$P_H$ measured, reconstructed & $0.5368$ & $0.6455$ & $0.3210$ \\
$P_H$ predicted, sudden       & $0.5526$ & $0.6703$ & $0.3185$ \\
$P_H$ measured, raw           & $0.0027$ & $0.0011$ & --- \\
\bottomrule
\end{tabular}
\caption{The sudden formula against the propagated result, over all $12\,878$ rays of M15-7b
at $t = 3.00$~s and $E = 20$~MeV.}
\label{tab:subita}
\end{table}

The Pearson correlation between predicted and measured is $0.9987$ and the Spearman $0.9929$;
the median ratio is $0.976$ with a $10$--$90\%$ range of $0.850$ to $1.007$; and $98.0\%$ of
rays agree to better than $0.05$ in absolute $P_H$, $99.8\%$ to better than $0.10$. The raw
profile returns $P_H < 0.05$ in $99.9\%$ of the rays. The relative shortfall shrinks as the
predicted jump grows: the median ratio rises from $0.897$ for predicted $P_H < 0.2$ to $0.991$
for predicted $P_H > 0.8$ (Spearman $\rho = +0.50$ over the $12\,845$ rays whose prediction is
non-zero). This is a consistency test of the calculation for the front that was placed: it
shows that the reconstructed profile, a $2.2$~km transition, behaves close to a sudden
crossing between the two densities, not that the front was placed where the physical one
is.

\section{Where the comparison is relevant}
\label{sec:quando}

The difference of Table~\ref{tab:tresvias} is largest where the density jump spans the
resonance, and it needs the numerical front to be wide enough to suppress what a sudden
crossing would give.

\subsection{The band in energy}
\label{sec:banda}

Behind the front the density is the unshocked value times the compression $C$; ahead of it,
the unshocked value itself. Taking $Y_e$ equal on both sides, the energies whose resonance
density lies between the two run from
\begin{equation}
  E_{-} \;=\; \frac{15840\ \mathrm{MeV\,g\,cm^{-3}}}{Y_e\,C\,\rho_0(r_s)}
  \qquad\text{to}\qquad
  E_{+} \;=\; \frac{15840\ \mathrm{MeV\,g\,cm^{-3}}}{Y_e\,\rho_0(r_s)},
  \label{eq:banda}
\end{equation}
with $\rho_0$ the presupernova density at the shock radius $r_s$, so that the band spans a
factor $C$ and its location is set by $Y_e\rho_0$ at the front. The sudden formula does not
vanish beyond these edges: with both densities on the same side of the resonance the two
matter angles still differ, so $P_H$ has tails, as Fogli \emph{et al.}~\cite{fogli2003} note.
Evaluated at $C = 5.25$, the band approximates the interval above half the peak: the
half-maximum points lie $8\%$ above the lower edge in energy and $2\%$ above the upper one,
while $P_H$ falls to a tenth of the peak within a factor $1.4$ in energy of either edge, to a
hundredth within a factor $2.4$, and to $0.5\%$ at three times the upper edge.

Taking $\rho_0$ and $Y_e$ from the published presupernova models \citep{utrobin2019} and $r_s$
and $C$ from the simulation, Eq.~\eqref{eq:banda} reproduces the band edges measured ray by ray
in the three-dimensional output:

\begin{table}[h]
\centering
\begin{tabular}{llcc}
\toprule
model & $t$ (s) & from Eq.~\eqref{eq:banda} (MeV) & measured in 3D (MeV) \\
\midrule
W18-3  & $1.68$ & $1.5$ -- $6.8$    & $1.4$ -- $6.6$ \\
W18r-2 & $1.69$ & $2.0$ -- $10.0$   & $1.9$ -- $10.0$ \\
W20-5  & $1.69$ & $5.6$ -- $28.2$   & $5.6$ -- $28.3$ \\
W16-3  & $1.69$ & $5.3$ -- $24.7$   & $5.3$ -- $30.8$ \\
W18x-2 & $1.68$ & $11.7$ -- $62.6$  & $11.3$ -- $61.6$ \\
W18-3  & $8.72$ & $38.9$ -- $185.0$ & $36.3$ -- $176.4$ \\
W18r-2 & $8.67$ & $47.6$ -- $233.5$ & $44.8$ -- $222.8$ \\
W20-5  & $8.67$ & $46.9$ -- $225.0$ & $45.5$ -- $222.7$ \\
W16-3  & $8.70$ & $55.3$ -- $265.3$ & $54.1$ -- $269.1$ \\
W18x-2 & $8.71$ & $94.0$ -- $473.9$ & $87.5$ -- $456.9$ \\
\bottomrule
\end{tabular}
\caption{Band edges from Eq.~\eqref{eq:banda} against those measured on the three-dimensional
models, $24$ rays per entry, ordered by the lower edge. Over the twenty edges the median
discrepancy is $3.6\%$ and nineteen are below $10\%$; the exception is the upper edge of W16-3
at $1.69$~s, at $20\%$. The comparison is of band edges inferred from densities, not of $P_H$
or of event rates.}
\label{tab:banda}
\end{table}

Equation~\eqref{eq:banda} is a geometric estimator: it avoids a flavour calculation, but it
needs the shock radius and the compression, which here come from the hydrodynamics. Its use
is to locate where the effect can be large for a given progenitor and shock trajectory. One
consequence matters for comparisons between models. The five W progenitors sit at the same
epoch to within $10$~ms, and their bands there run from $1.4$--$6.6$ to $11.3$--$61.6$~MeV, a
factor of $43$ from the lowest edge to the highest, with no single energy inside all five. A
comparison at one fixed energy may therefore miss the band of some of the models: at
$10$~MeV, inside W16-3's band at $1.69$~s but above W18-3's, W18-3 returns
$\Pee = 0.0922 \pm 0.0043$ raw against $0.0944 \pm 0.0109$ reconstructed, a difference of
$0.19\sigma$, because at that energy its resonance lies $4740$~km ahead of the front, in
unshocked material. In the configurations of Table~\ref{tab:tresvias} the resonance sits on
the front, with median offsets of $0$ and $+230$~km.

The radii at which the band is detectable, $3\times10^{4}$ to $1.5\times10^{5}$~km, lie inside
the helium core, whose boundary in these progenitors is at $4$ to $5\times10^{5}$~km. The
density the band probes is therefore set by a part of the star that the colour of the envelope
does not determine. That does not establish that the shock trajectory, the compression or the
reverse-shock structure carry over to a red supergiant, and those enter
Eq.~\eqref{eq:banda} too. As a bracket from progenitor structure only: against the
red-supergiant progenitors of Sukhbold, Woosley \& Heger~\cite{sukhbold2018}, whose radii are
$880$ to $1130\,R_\odot$ against $29$ to $56\,R_\odot$ here, the shock radius at which the band
contains a given energy agrees to between $8$ and $22\%$, and the unshocked density at
$1.1\times10^{5}$~km is $92$ to $130$~g\,cm$^{-3}$ against $95$ to $157$. This uses the
reduced-mass-loss subset of that grid, and the compression $C = 4.7$ of the blue-supergiant
models.

\subsection{Energy and epoch}
\label{sec:energia}

At a fixed epoch the resonance moves outward as the energy rises, since
$\rho_{\rm res} \propto 1/E$. In M15-7b at $t = 3$~s it sits on the front at $5$ to $20$~MeV,
$r_{\rm res}/r_{\rm front} = 1.01$--$1.02$, and ahead of it at $40$ and $60$~MeV, $1.26$ and
$1.44$ (Appendix~\ref{ap:energia}). The raw profile stays within $3\%$ of the adiabatic value
at all five energies, although $w/\Losc$ falls with $E$; the reconstructed one exceeds it by a
factor of $6$ to $10$ where the resonance is on the front and by $2.30$ and $1.18$ at $40$ and
$60$~MeV. The change is largest where the resonance lies within the jump; ahead of it a
smaller change remains, as the tails of the sudden formula lead one to expect, and the width
decides how much of the available effect survives \citep{fogli2003}.

Shock features are therefore visible ``during a limited time window'' as they migrate through
the detectable range \citep{fogli2003,tomas2004}. Evaluated with one density read at the front,
\begin{equation}
  E_{\rm res}(t) \;=\; 20\ \mathrm{MeV} \times \frac{792}{(Y_e\rho)_{\rm front}(t)},
  \label{eq:eres}
\end{equation}
lies between the edges of Eq.~\eqref{eq:banda} and marks where the band is, not its centre. In
M15-7b
$E_{\rm res}$ is $6.3$~MeV at $t = 3$~s and $2.1$~GeV at $t = 10$~s. In W16-3 at $8.70$~s it is
$66$~MeV, still in the detectable range, and reconstruction there gives $P_H = 0.526$
(Table~\ref{tab:tresvias}): the relevant energy window persists in one model after it has disappeared in another, so when shock effects end is a property of the progenitor. The window closes when the
band, with its tails, leaves the energies that the source emits and the detector records.

M15-7b provides a control at its second usable epoch, $t = 10$~s, when the front, at
$1.44\times10^{5}$~km, has overtaken the resonance, at $1.38\times10^{4}$~km, by a factor of
ten. There the raw and reconstructed profiles give $\Pee = 1.026$ and $1.027$ times the
adiabatic value, medians over rays, with a ray-to-ray scatter of $7\%$; at $3$~s, with the
front on the resonance ($3.40$ against $3.45\times10^{4}$~km), the reconstructed median is
$7.27$ times the adiabatic value, with a scatter of $56\%$. The procedure that raises $P_H$ by
about half when the front crosses the resonance leaves it unchanged when it does not. At
$20$~MeV the two epochs bound the disappearance of the effect in this model between $3$ and
$10$~s, without locating it within that interval; at $5$~MeV a median $P_H$ of $0.019$ remains
at $10$~s (Table~\ref{tab:orcamento}).

\subsection{Width of the front}
\label{sec:criterio}

The crossing probability depends on the density scale height at the resonance through
$\gamma_{\rm res} \propto H$, so the width of the front controls it. Imposing a front of
controlled width $w$ on the parametrised profile of Section~\ref{sec:mapa} and scanning it:

\begin{table}[h]
\centering
\begin{tabular}{rrl}
\toprule
$w$ (km) & $w/\Losc$ & $\Pee$ \\
\midrule
$200$ & $3.0$   & $0.0229$ \\
$60$  & $0.90$  & $0.0907$ \\
$20$  & $0.30$  & $0.0617$ \\
$6$   & $0.09$  & $0.1460$ \\
$2$   & $0.03$  & $0.1717$ \\
$0.6$ & $0.009$ & $0.1306$ \\
\bottomrule
\end{tabular}
\caption{Dependence on the width of the front, at $t = 1$~s and $E = 20$~MeV, on the
parametrised profile; the adiabatic value is $\Pee = \sthirteen = 0.0222$.}
\label{tab:largura}
\end{table}

At $3\,\Losc$ the response is close to the adiabatic value --- $\Pee = 0.0229$ against
$0.0222$, which is $P_H \simeq 0.0025$, the size of what the raw M15-7b profiles return. Between
$0.9$ and $0.09\,\Losc$ it rises, non-monotonically, by a factor of six. Below $0.1\,\Losc$ the
three widths sampled give $0.146$, $0.172$ and $0.131$: the response no longer rises
systematically, but it varies by $\pm15\%$, which may be interference between the two edges of
the shell, $3000$~km apart, if coherence holds across them (Section~\ref{sec:evolucao}); we have
not separated the limit of each edge from that interference, and these three values do not
establish convergence. On a ray of M15-7b the same scan, with the $\tanh$ width $w$ of
Table~\ref{tab:protocolo}, gives $P_H$ still rising between $w = 1$~km and $0.1$~km, $0.015$ to
$0.0015\,\Losc$. In the twelve configurations of Table~\ref{tab:tresvias} the reconstructed
values reach their narrow limit by $w = 0.1$~km, and the $w = 1$~km values used throughout
differ from it by $-0.042$ to $+0.055$ (Section~\ref{sec:consequencia}), most at $5$~MeV.

The figure of $3\,\Losc$ is a property of this parametrised experiment and not a hard cutoff:
a front of finite width has a Fourier distribution rather than a single scale, and in a
composite field the long modes act through the distorted-phase channel \citep{patton2014}. We
use it only as a reference. The stored widths of Section~\ref{sec:modelos} are $3.5$ to
$4.7\,\Losc$ at the $20$~MeV reference and $4.2$ to $14.1\,\Losc$ at the energies of the
early-epoch configurations; the late-epoch widths were not measured. The angle-averaged
profiles, at $37$ to $673\,\Losc(20~\mathrm{MeV})$, are far wider.

\subsection{Crossing probability over time and energy in one model}
\label{sec:orcamento}

Table~\ref{tab:orcamento} collects the reconstructed $P_H$ of M15-7b at the two usable epochs
and five energies, with $P_H$ obtained by inverting $\Pee$ (Section~\ref{sec:ph}). It
characterises one model and is not an uncertainty budget for an analysis, which would need a
representative progenitor set and the joint distribution of $P_H(t,E)$ along each line of
sight; entries along one line of sight are correlated across energy and epoch.

\begin{table}[h]
\centering
\input{ph_tabela.tex}
\caption{Effective crossing probability $P_H$ from M15-7b, from the inversion of $\Pee$:
medians over $20$ rays for the raw and the reconstructed profile, and the $10$--$90\%$ range of
the reconstructed one across rays. Near the edges of the band the response is bimodal across
rays --- the resonance falls inside the front for some and outside for others --- so the
median there is poorly determined.}
\label{tab:orcamento}
\end{table}

The largest median is $P_H = 0.808$ at $t = 3$~s and $10$~MeV, with a $10$--$90\%$ range of
$0.74$ to $0.84$, closer to fully non-adiabatic than to the adiabatic limit. Towards high energy
the median drops to $0.004$ at $60$~MeV, with a range of $0.00$ to $0.04$; the two ranges do not
overlap. We do not interpret the lower median at $20$~MeV, $0.48$, whose range runs from $0.08$
to $0.85$: the response there is bimodal and the median of $20$ rays is poorly determined. By
$10$~s the median has fallen to $0.002$ or below at every energy but $5$~MeV, where $0.019$
remains with a range reaching $0.12$.

\subsection{Dispersion and the uncertainty of the reconstruction}
\label{sec:visada}

That transition probabilities are ``random variates unique to each ray'' is known
\citep{knellermauney2013}. The $38\%$ ray-to-ray spread in the front radius of M15-7b is the
same phenomenon driven by shock asphericity, and it propagates into the conversion strongly:
across the $40$ rays of Figure~\ref{fig:raios}, $m_3$ ranges from $0.108$ to $0.9999$ and
$\Pee/\sthirteen$ from $1.0$ to $12.0$, with a scatter of $56\%$ about the mean, against $4\%$
for the raw profiles.

Part of that spread is the reconstruction. The two front detectors disagree on where the front
is by about one grid zone --- at the resonance radius $177$~km, or $2.7\,\Losc(20~\mathrm{MeV})$.
The velocity detector decides which rays have a front and supplies the radius we report; the
reconstructed front is placed by the density detector. Propagating the same $24$ rays twice,
with the reconstructed front placed by each, the paired difference in $P_H$ is $-0.012 \pm 0.068$: no statistically significant systematic shift is found in that configuration, and with $24$ rays a shift smaller than about a third of the effect would
not be; the other eleven configurations were not tested this way. Ray by ray the difference
reaches $0.78$, with a scatter of $0.33$. The distribution between rays therefore combines
physical dispersion with the uncertainty of the reconstruction, and cannot be read in full as
what a set of observers would see. A supernova is observed along one line of sight, so a single
event is one draw from such a distribution, not its mean.

\section{Smoother structures and turbulence}
\label{sec:estruturas}

The paired comparison does not require that other structures be irrelevant, but whether they
could compete is a natural question. Appendix~\ref{ap:perturbativo} reports perturbative tests
on a parametrised profile with a single resonance crossing, imposing weak density modes one at
a time and as broadband fields. Their content, with the assumptions each rests on:

\begin{table}[h]
\centering
\small
\begin{tabular}{@{}>{\raggedright\arraybackslash}p{0.22\textwidth}>{\raggedright\arraybackslash}p{0.33\textwidth}>{\raggedright\arraybackslash}p{0.37\textwidth}@{}}
\toprule
structure & what was measured & what the result assumes \\
\midrule
single weak mode, wavelength $\lambda$ & response peaks just below $\Losc$, falls as $\lambda^{1.4}$ below it, and is at the $10^{-7}$ floor above it & one mode at a time, contrast $3\times10^{-3}$, one crossing, incoherent observable \\
turbulence below $\Losc$ & leakage $\propto \sigma^{2.02 \pm 0.03}$; at the extrapolated contrast, a median $3.9\times10^{-3}$ of $\Pee$, at most $2.0\times10^{-2}$ across models & sub-grid contrast extrapolated with a measured slope; one crossing; no combination with multiple resonances \\
large-scale flows (convection, SASI) & not tested directly; single long modes are at the floor & combinations of modes, displaced resonances and changed jumps not tested \\
shock front & Sections~\ref{sec:3d}--\ref{sec:quando} & treated by propagation and the sudden formula, not by the weak-mode law \\
\bottomrule
\end{tabular}
\caption{What the perturbative tests of Appendix~\ref{ap:perturbativo} establish, and under
which assumptions.}
\label{tab:criterio}
\end{table}

Two points bound how these results can be used. The weak-mode response is a statement about
the incoherent observable for one mode at a time; in composite fields long modes can suppress
transitions driven by others through the distorted-phase channel \citep{patton2014}, and
several resonances can raise the sensitivity to small amplitudes \citep{kneller2010}, neither
of which these tests include. And the quadratic law describes the weak regime: a strong jump is
described by the sudden formula of Section~\ref{sec:subita}, not by extending the law. Under the
stated assumptions, turbulence at the extrapolated contrast is small at the H resonance, which
agrees with the conclusion Borriello \emph{et al.}~\cite{borriello2013} reached by propagating
real profiles; the amplitude at the scales that act is below the grid of every model available,
so that conclusion inherits the extrapolation.

\section{Implications for DUNE: a parametrised demonstration}
\label{sec:dune}

Whether a change in $P_H$ moves the event count, and in which direction, depends on the
flavour composition of the flux, which we have not folded with the three-dimensional fronts of
Section~\ref{sec:3d}. This section is a separate calculation: a parametrised progenitor and
front, propagated over time and energy and folded with a published emission model. It shows
the size and the structure of the consequence for a front of the kind we impose.

\subsection{Where a constant prescription fails}
\label{sec:mapa}

Onto the power law $\bar\rho(r) = 10^{10}\,(100\,\mathrm{km}/r)^3$~g\,cm$^{-3}$ we impose a
propagating front in the spirit of Schirato \& Fuller~\cite{schirato2002} and Fogli
\emph{et al.}~\cite{fogli2003}: ahead of $r_f(t) = v_f t$ the progenitor profile, behind it a
compressed shell, and inside that the rarefied region left by the expansion. The nominal
parameters are in Table~\ref{tab:protocolo}; sampling of the front is refined to
$\simeq 8$~km, against $\Losc = 67$~km.

\begin{figure}[t]
\centering
\includegraphics[width=0.82\textwidth]{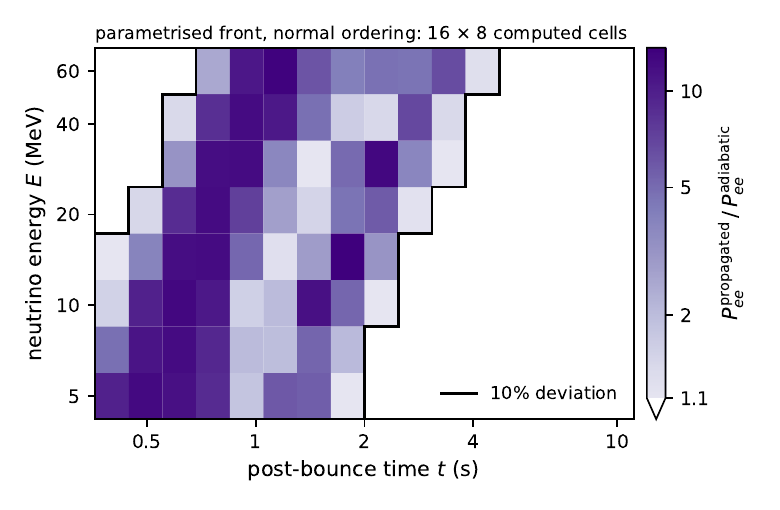}
\caption{Ratio of the propagated $\Pee$ to the adiabatic constant, over post-bounce time and
neutrino energy, for the \emph{parametrised} front in normal ordering. Each cell is one
computed point, $16$ instants by $8$ energies, with nothing interpolated between them; white
cells deviate by at most $10\%$, and the outline encloses those that deviate by more. The band
is diagonal because the resonance radius scales as $E^{1/3}$, so the shock reaches it later
for higher energy.}
\label{fig:mapa}
\end{figure}

The diagonal band of Figure~\ref{fig:mapa} is the migration described by Fogli
\emph{et al.}~\cite{fogli2003}. The peak error in the survival probability is a factor $13.7$
at $t = 1.8$~s and $E = 14.5$~MeV; the cells in which the prescription is wrong by more than
$10\%$ cover $55\%$ of the sampled plane and end at $t = 4.2$~s; outside them the ratio is
$1.00$. The onsets grow by a factor $1.39$ between $21$ and $60$~MeV, against $1.43$ for
$t \propto E^{1/3}$. Doubling the time sampling from $16$ to $40$ instants changes the peak
ratio from $13.7$ to $13.5$, the fraction above $10\%$ from $55\%$ to $56\%$ and the closing
time from $4.2$ to $4.0$~s, while the number of local maxima per energy rises from $2$--$3$ to
$5$--$11$: these tests indicate numerical convergence of the overall envelope, although the finer
temporal structure within it remains unresolved. Varying
the front velocity from $10^4$ to $4\times10^4$~km\,s$^{-1}$, the compression from $4$ to $10$
and the rarefaction from $0.05$ to $0.30$, one at a time, keeps the peak ratio within
$11.9$--$13.6$ and moves the event-count bias below between $-3.5\%$ and $-6.7\%$. The front
velocity measured in M15-7b, $1.15\times10^{4}$~km\,s$^{-1}$, and its compression over the
density bracket, $5.22$, fall inside those ranges; over the velocity bracket the compression,
$3.7$, lies just below. The location of the band does not transfer: the parametrised map
places the affected energies higher at $t = 3$~s than M15-7b does, because its front is
faster and its background a power law.

\subsection{Event counts}
\label{sec:contagens}

Folding the Nakazato \emph{et al.}~\cite{nakazato2013} emission ($20\,M_\odot$, revival time
$100$~ms, metallicity $0.004$, Shen equation of state) through {\sc SNOwGLoBES} for the
\texttt{ar40kt} configuration at $10$~kpc, summed over $t \in [0.4, 10]$~s and
$E \in [5, 60]$~MeV, the $\nue$ charged-current count falls from $1241.8$ with the adiabatic
constant to $1187.0$ with the propagated conversion, a bias of $-4.4\%$. The flux is sampled on
$48\times24$ points in $(t,E)$ and $P_H$ is propagated on a $32\times16$ grid and interpolated
onto it. Refining the flux sampling from $16\times8$ moves the counts by $5.9\%$ and the bias by
less than a tenth of a percentage point; refining the $P_H$ grid moves the bias from $-4.1\%$
to $-4.4\%$; between the two finest samplings the counts still move by $1.2\%$ and the bias by
$0.05$ percentage points. The bias is therefore the quantity we report.

\begin{figure}[t]
\centering
\includegraphics[width=0.82\textwidth]{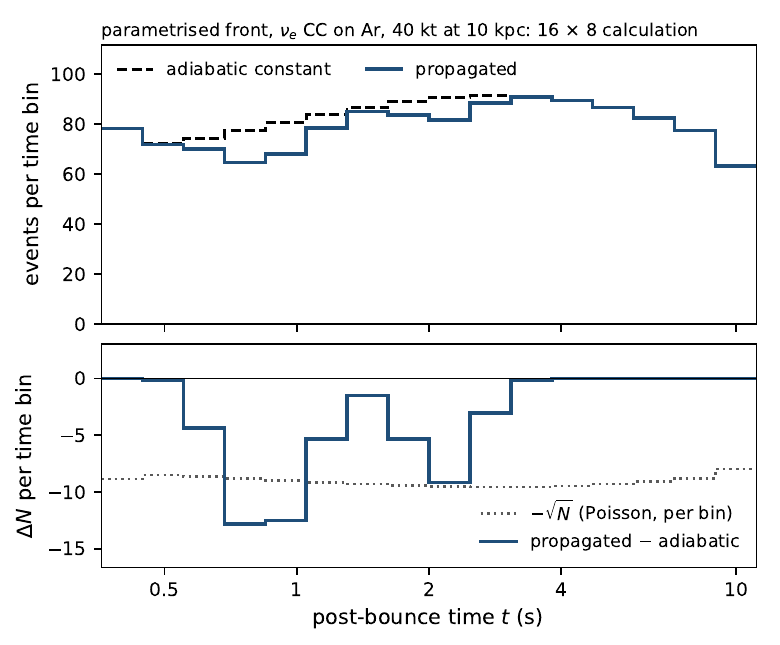}
\caption{$\nue$ charged-current events in a $40$~kt argon detector at $10$~kpc against
post-bounce time, summed over detected energy, with the adiabatic constant and with the
conversion propagated through the parametrised front. \emph{Bottom:} the difference per time
bin, against the Poisson fluctuation of each bin. This is the $16\times8$ calculation from
which the per-bin separations come, with totals $1314.7$ and $1260.3$ ($-4.1\%$); the totals
quoted in the text are from the finer grid, which has no per-bin output.}
\label{fig:eventos}
\end{figure}

As a measure of the separation between the two predictions under Poisson statistics we use
$S = [\sum_b (\Delta N_b)^2/N_b]^{1/2}$ over bins $b$, with $N_b$ the adiabatic count. On the
fine-grid totals $S = 1.6$. On the $16\times8$ calculation of Figure~\ref{fig:eventos} it is
$2.7$ over the $(t,E)$ bins, $2.5$ projected on time and $1.6$ projected on energy: the deficit
is confined to $0.5$ to $3.4$~s, no single time bin departs by more than $1.5$ times its Poisson
fluctuation, and the separation comes from the accumulation of same-sign deviations over many
bins. For a fixed detector and source $S$ scales as $1/d$, and reaches $13$ at $2$~kpc on the
coarse grid. These are distances between two fixed templates, not significances after
marginalising over the source and the detector response, and we have not computed the latter.

The sign of the change is set by the source. Writing the transformed flux as
$\Pee F_{\nue} + (1-\Pee)F_{\nu_x}$, the shift is
\begin{equation}
  \Delta N \;\propto\; \sum_{t,E} \sigma(E)\,
  \Delta \Pee(t,E)\, \bigl[F_{\nue}(t,E) - F_{\nu_x}(t,E)\bigr],
  \label{eq:sinal}
\end{equation}
so its sign follows the flavour hierarchy of the source inside the band, and a larger $\Pee$
alone does not fix the sign or the size of the change in events. Across four families the
magnitude is a few per cent and the sign is not universal: Nakazato \citep{nakazato2013}
gives $-4.0$ to $-4.2\%$, the Garching models \citep{mirizzi2016} $-3.9$ and $-4.4\%$,
Sukhbold \citep{sukhbold2016} $-2.9\%$, and Fornax~2022 \citep{wang2022} $+1.3$ and $+1.4\%$.
Equation~\eqref{eq:sinal} with $\sigma \propto E^2$ reproduces the sign in all six cases.
Because the sign follows the source, no single correction factor removes the bias for all
sources, and it is degenerate with the spectral quantities being extracted.

We also examine both neutrino mass orderings. In normal ordering the H resonance occurs in the
neutrino sector, whereas in inverted ordering it occurs in the antineutrino sector, and the
peak ratios follow: $13.7$ and $1.0$ for
$\nue$ and $\bar\nu_e$ in normal ordering, $2.3$ and $30.4$ in inverted. Argon is a $\nue$
target: the $\bar\nu_e$ channel carries $17$--$33$ events against $1000$--$1300$, so the factor
$30$ is not usable. The bias is $-4.4\%$ in normal ordering and $-5.4\%$ in inverted; the
latter was computed on the coarser $P_H$ grid and has not been through a convergence study of
its own.

The $2.3$ in the $\nue$ channel for inverted ordering is not H contamination: at the instant
the front reaches the H resonance of each energy --- $0.58$ to $1.33$~s for $5$ to $60$~MeV on
this profile --- the $\nue$ ratio is $1.001$ in the median while the $\bar\nu_e$ ratio is
$20.5$. The timing points to the L resonance. Resonance densities scale as
$\Delta m^2 \cos 2\theta/E$, so on the $r^{-3}$ background the L resonance lies further out than
the H resonance by
$[(\dmsq \cos 2\theta_{13})/(\Delta m^2_{21} \cos 2\theta_{12})]^{1/3} \simeq 4.35$ in radius,
and the front, moving at constant speed, reaches it $4.35$ times later: at $2.5$~s for $5$~MeV
and $5.8$~s for $60$~MeV. The centroid in time of the $\nue$ excess is $2.8$~s at $5$~MeV and
$6.0$~s at $60$~MeV (Spearman $\rho = +0.93$ over energy, $p = 9\times10^{-4}$), close to the L
crossings and far from the H ones. This estimate takes $Y_e$ equal at the two radii and
two-flavour resonance conditions; we have not computed the crossings from the full
three-flavour condition.

\subsection{Consequences for propagators and analyses}
\label{sec:pratica}

The development branch of {\sc SNEWPY} (commit \texttt{95a9eed}, October~2026; the latest
release, 1.6, dates from April~2025) adds an \texttt{MSWEffect} transformation which propagates ``a density profile
and electron fraction provided by the user'' through {\sc SNOSHEWS}.\footnote{\url{https://github.com/SNEWS2/SNOSHEWS};
the module is by BenZvi, Colomer Molla, Habig, Kneller, Migenda, Scholberg, Sheshukov \& Tseng.
What we state here about its treatment of profiles is read from its source and example data
at commit \texttt{9246c19} (July~2025), the latest as of October~2026.} It splits its integration into domains at discontinuities, but
recognises a discontinuity only where two consecutive radii of the input file are equal, as in
its own example profile; a front from a hydrodynamical simulation is a monotone ramp across
$1.6$--$1.7$ zones, which an adaptive integrator resolves to whatever accuracy is demanded. It
can then return an artificially adiabatic evolution, as the raw profiles of
Table~\ref{tab:tresvias} do where the width suppresses the transition, though not completely in
every case (raw $P_H$ up to $0.215$). Hydrodynamic resolution and integrator precision are
therefore separate questions: tightening the tolerance on a stored profile converges on the
answer for the stored ramp, and only a reconstruction of the front addresses the first.

For analyses, the effect concerns the signal after the shock reaches the resonance ---
time-dependent fits and spectral parameters integrated over the burst --- and not the
neutronization burst at $t \lesssim 25$~ms, which is the primary handle in Sarmah
\emph{et al.}~\cite{ordering2026}. That $P_H \to 1$ in normal ordering reproduces the inverted
adiabatic value is an identity of the standard formulae; a table such as
Table~\ref{tab:orcamento} does not resolve that degeneracy, but it is the kind of input a fit
needs in order to marginalise over it. Whether the effect could be inferred rather than
marginalised --- from its location in $(t,E)$, which Eq.~\eqref{eq:banda} relates to
$Y_e\rho_0$ at the shock radius --- depends on the source model, the detector and their
degeneracies, which we have not analysed.

\section{Limitations}
\label{sec:limites}

\paragraph{The reconstruction is a prescription.} We replace the numerical front by a
$\tanh$ transition of width parameter $w = 1$~km ($2.2$~km from $10$ to $90\%$), preserving the
densities on both sides, placed by a density criterion about one zone from the velocity
discontinuity (Table~\ref{tab:protocolo}). It approximates a sudden crossing without being one:
narrowing it to $w = 0.03$~km changes the differences of Table~\ref{tab:tresvias} by up to
$0.055$, most at low energy, where the residual width is a larger fraction of $\Losc$. The width
scan of the parametrised front (Section~\ref{sec:criterio}) does not establish a converged
observable below $0.1\,\Losc$, and a resolved calculation of the flow on the physical scale of
the shock would be needed to do better. The reported differences are therefore a property of this procedure.

\paragraph{Estimators and uncertainties.} The entries of Table~\ref{tab:tresvias} are means over
$20$ rays, with standard errors from a rerun that records every ray; the standard error treats
the rays as independent draws, which neighbouring cells are not. The detector-placement test
was run in one configuration. Table~\ref{tab:orcamento} reports medians and ranges, not uncertainties on the medians, because
the per-ray values of that run were not stored. The control values at $3$ and $10$~s of
Section~\ref{sec:energia} were obtained in an earlier analysis for which the original script is
not included with the other analysis scripts; they can be checked only by repeating that
propagation.

\paragraph{Angular sampling.} The $12\,878$ rays are grid cells, not a solid-angle-weighted
sample of the sky. Each Yin-Yang patch is uniform in $(\theta,\varphi)$ over a band of
colatitude, so the solid angle per cell varies as $\sin\theta$ across it: $137$ cells over the
$270^\circ$ of the patch in $\varphi$ gives $1.97^\circ$ per cell, and $47$ cells of that size
in $\theta$ span $92.6^\circ$, the nominal $90^\circ$ band \citep{kageyama2004} plus the overlap
at each end. Cells at the band edges subtend $\sin 45^\circ = 0.71$ of those at its centre, the
overlap represents some directions twice, and neighbouring rays are correlated. Per-ray
averages therefore describe the set of cells and not the distribution a randomly oriented
observer would sample, and $s/\sqrt{N}$ over cells is a formal standard error under
independence, not the uncertainty of an angular average.

\paragraph{Coherence.} The evolution is coherent by assumption, and retains interference
between features far apart --- the edges of the parametrised shell, $3000$~km apart at
$t = 1$~s, and distant crossings in the three-dimensional profiles --- that decoherence could
reduce (Section~\ref{sec:evolucao}). Structure we attribute to interference is conditional on
it.

\paragraph{Six models, two epochs each, all blue supergiants.} The six progenitors are all but
one of the blue-supergiant models the archive distributes, setting aside the two hand-made
progenitors its curator advises against; the exception is M158b-1, which we have not obtained.
Of the four epochs distributed, only those at $3$ and $10$~s (M15-7b) or $1.7$ and $8.7$~s (the
W models) contain the H resonance within their domain, so the time dependence is coarse. The
next galactic supernova may have a red-supergiant progenitor; Section~\ref{sec:banda} gives a
bracket from progenitor structure, not a calculation, and $P_H$ for a red supergiant needs the
simulations, since the raw term depends on the numerical front of a run we do not have.

\paragraph{The detector calculation.} Section~\ref{sec:dune} uses a parametrised front and an
emission model from another calculation, charged-current absorption on argon only, and Poisson
statistics, with no detector background, no reconstruction systematics beyond what
{\sc SNOwGLoBES} applies, and no marginalisation over astrophysical parameters.
Neutrino--neutrino refraction is outside the framework.

\section{Conclusions}
\label{sec:conc}

\begin{enumerate}
\item \textbf{In the procedure studied, reconstructing the front changes the computed
conversion substantially.} Across twelve configurations of six three-dimensional
blue-supergiant models, each chosen inside the band in which the front meets the resonance,
propagating the same rays with the front replaced by a $2.2$~km transition raises the
ray-averaged crossing probability by $0.44$ to $0.76$ in absolute terms, median $0.56$, with
standard errors of $0.025$ to $0.076$; narrowing the transition to $0.07$~km moves each value by
at most $0.055$, to a range of $0.41$ to $0.79$. The stored fronts are $1.6$ to $1.7$ zones
wide in every model. Along all rays of one
configuration the sudden formula, evaluated on each ray's densities, agrees closely with the
propagated reconstructed result.

\item \textbf{The change is largest where the density jump spans the resonance.} That happens
over a band of energies that Eq.~\eqref{eq:banda} estimates from the presupernova density,
given the shock radius and the compression, to a median $3.6\%$ in the band edges, with tails
outside it; bands of sibling progenitors at one epoch share no energy, so comparisons at a
fixed energy can miss some of them. The window is progenitor-dependent. The result also depends
on the reconstruction: its residual width was tested in one model, the response on a
parametrised front is still varying below $0.1\,\Losc$, and the placement of the front is
uncertain by about one zone. Hydrodynamic resolution and integrator
precision are distinct, and tightening the tolerance on a stored profile does not substitute
for reconstructing the front.

\item \textbf{The consequence for DUNE is of order per cent in a parametrised demonstration,
and remains to be computed consistently.} With a parametrised front and a published emission
model, time-dependent conversion changes the $\nue$ charged-current count in a $40$~kt argon
detector by $-4.4\%$, with a sign that follows the flavour composition of the source and a
comparable size in both orderings. Folding the reconstructed three-dimensional fronts with an
emission model from the same simulations is the calculation this motivates.
\end{enumerate}

Perturbative tests on a profile with one crossing find, under the assumptions of
Table~\ref{tab:criterio}, that weak modes longer than $\Losc$ do not act alone and that
turbulence at the extrapolated sub-grid contrast is small; these are conditional statements and
the comparison above does not depend on them. The practical recommendation is that a flavour
transformation which propagates through a hydrodynamical profile should reconstruct the front
before computing the evolution.

\section*{Acknowledgements}

We are grateful to H.-Th. Janka for granting R.C.R.L. access to the Garching Core-Collapse
Supernova Archive and for several discussions that shaped this work: on which progenitor
models are suitable and which are not, and on the numerical nature of the shock width, which
became Section~\ref{sec:3d}. We thank D. Kresse for arranging the access. This work made
use of the Garching Core-Collapse Supernova Archive,
\url{https://wwwmpa.mpa-garching.mpg.de/ccsnarchive/}.

\section*{Data and reproducibility}

Computations use {\sc GHOST} \citep{ghost}, a three-flavour propagator written for this work
and publicly available at \url{https://github.com/RafaelCRdeLima/GHOST} under the MIT licence,
validated against {\sc NuOscProbExact} \citep{bustamante2019}; each output records a hash of
the numerical core that produced it. Event rates use {\sc SNOwGLoBES} \citep{snowglobes} through
{\sc SNEWPY} \citep{snewpy2021}. The supernova model data --- both the three-dimensional
Yin-Yang output and the angle-averaged radial profiles --- are from the Garching
Core-Collapse Supernova Archive, model M15-7b of A.~Wongwathanarat
\citep{wongwathanarat2015,utrobin2015,utrobin2019,utrobin2021}, and the five blue-supergiant
explosions W16-3, W18-3, W18r-2, W18x-2 and W20-5 of the same archive \citep{utrobin2019},
used with permission and not redistributed here. The public inputs are fetched by the
scripts that use them: the presupernova structures of Utrobin
\emph{et al.}~\cite{utrobin2019} from the CDS (J/A+A/624/A116), the red-supergiant progenitors
of Sukhbold, Woosley and Heger~\cite{sukhbold2018} from the Harvard Dataverse, and the
solar-neutrino survival probabilities of Borexino~\cite{borexino2021}. The analysis scripts
that produce the tables and figures, with the exception noted in Section~\ref{sec:limites}, and
the per-ray results of the twelve configurations of Table~\ref{tab:tresvias} at every
reconstruction width, recorded with the hash of the numerical core and of the scripts that
produced them, are not part of the public repository and are available from the authors on
request.

\addtocontents{toc}{\protect\setcounter{tocdepth}{1}}
\appendix

\section{Perturbative response to density structure}
\label{ap:perturbativo}

\subsection{The fluctuation field}
\label{sec:campo}

Where a stochastic perturbation is imposed we write $\rho(r) = \bar\rho(r)[1+\delta(r)]$ with
\begin{equation}
  \delta(r) \;=\; \sigma\,
  \frac{\sum_{n=n_{\min}}^{n_{\max}} a_n \cos\!\left(k_n r + \phi_n\right)}
       {\bigl[\sum_{n} \tfrac12 a_n^2\bigr]^{1/2}},
  \qquad
  a_n = n^{-p/2},\quad k_n = \frac{2\pi n}{L},\quad \phi_n \sim \mathcal{U}[0,2\pi),
  \label{eq:field}
\end{equation}
with $L = 100$~km, so that mode $n$ has wavelength $\lambda_n = L/n$, the power spectrum is
$\mathcal{P}(k_n) \propto n^{-p}$, and the normalisation enforces $\langle\delta^2\rangle =
\sigma^2$ exactly, independently of $p$ and $n_{\max}$. We quote the leakage
\begin{equation}
  \Delta \;\equiv\; \Pee - \sthirteen ,
  \label{eq:leakage}
\end{equation}
which vanishes in the adiabatic limit, and the noise-induced response
$R = \langle\Delta\rangle - \Delta_0$ with $\Delta_0$ the noise-free value at the same energy.
The profile is the $\rho \propto r^{-3}$ reference with a single H crossing.

\subsection{Scale}
\label{sec:escala}

In the stimulated-transition framework \citep{kneller2013sinusoidal,patton2014}, a density
mode of wavenumber $q_a$ drives transitions between matter eigenstates when
\begin{equation}
  \Bigl| \delta k_{12} + \sum_a n_a q_a \Bigr| \;\approx\; 0,
  \label{eq:parametric}
\end{equation}
so that the modes which act are those commensurate with the local splitting $\delta k_{12}$.
Setting $n_{\min} = n_{\max} = n$ in~\eqref{eq:field}, we measure the response to one mode at a
time at fixed $\sigma = 3\times10^{-3}$.

\begin{figure}[t]
\centering
\includegraphics[width=0.82\textwidth]{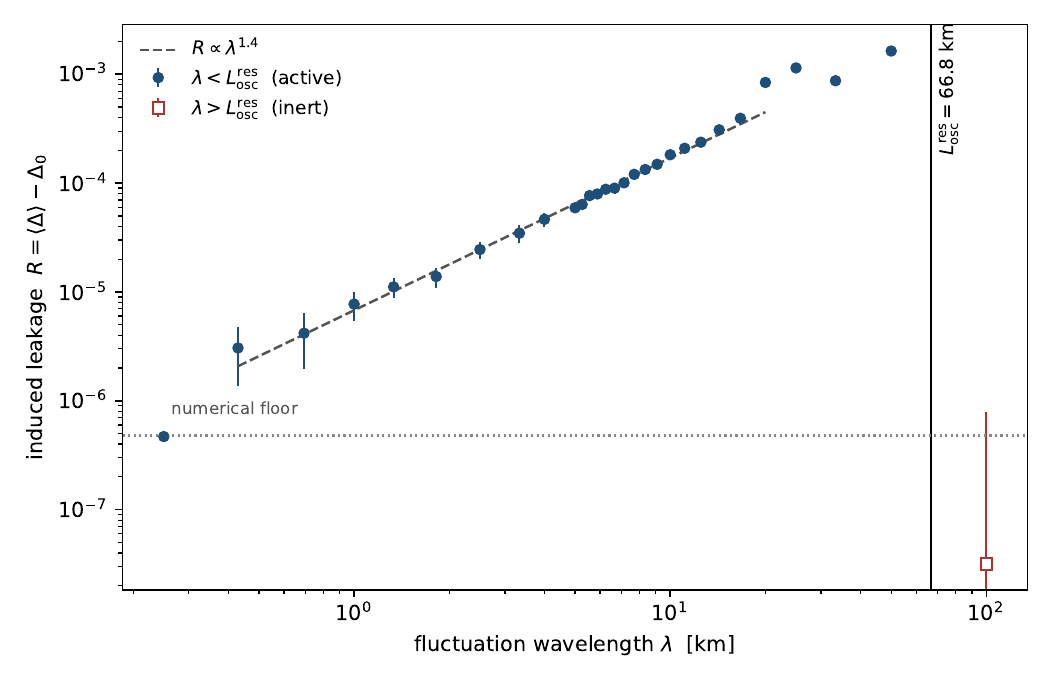}
\caption{Response to a single weak mode against its wavelength. The vertical line is
$\Losc = 66.8$~km; the dotted line is the numerical floor. In this test --- one mode at a
time, one crossing, incoherent observable --- modes near and below the resonant scale drive
transitions and modes above it do not.}
\label{fig:response}
\end{figure}

The single-mode response (Figure~\ref{fig:response}) peaks just below $\Losc$, is not monotonic
in its immediate neighbourhood --- $R$ at $25$~km exceeds $R$ at $33$ and $20$~km by ten
standard errors, which we attribute to the interference of the two matching points --- and
decays as $\lambda^{1.4\pm0.1}$ below it, reaching a numerical floor of $5\times10^{-7}$ near
$\lambda \simeq 0.5$~km. At $\lambda = 100$~km we measure $R = (3 \pm 75)\times10^{-8}$, five
orders of magnitude below the value at $50$~km. In the general case long-wavelength modes are
not inert: they enter through the distorted-phase channel, the factor $J_0(z_a)$ of Patton,
Kneller \& McLaughlin~\cite{patton2014}, which suppresses transitions that other modes drive.
When the long mode is alone there is no transition for it to suppress, since the smooth
crossing is adiabatic, so the floor is a statement about this observable and this
configuration. Holding the mode at $\lambda = L = 100$~km and varying the energy sweeps the
resonant scale past it at $E_c = L\,\dmsq \sin 2\theta_{13}/4\pi = 29.9$~MeV:

\begin{table}[h]
\centering
\begin{tabular}{rrcl}
\toprule
$E$ (MeV) & $\Losc$ (km) & $\lambda < \Losc$? & $\langle\Delta\rangle - \Delta_0$ \\
\midrule
 5 &  16.7 & no  & $(+1.1 \pm 1.7)\times10^{-7}$ \\
10 &  33.4 & no  & $(-1.28 \pm 0.30)\times10^{-7}$ \\
20 &  66.8 & no  & $(-4.5 \pm 6.0)\times10^{-7}$ \\
25 &  83.5 & no  & $(+5.3 \pm 2.0)\times10^{-7}$ \\
\midrule
30 & 100.2 & yes & $(+2.712 \pm 0.006)\times10^{-2}$ \\
35 & 116.9 & yes & $(+3.72 \pm 0.03)\times10^{-3}$ \\
40 & 133.6 & yes & $(+4.22 \pm 0.02)\times10^{-3}$ \\
60 & 200.5 & yes & $(+1.140 \pm 0.010)\times10^{-3}$ \\
80 & 267.3 & yes & $(+4.17 \pm 0.03)\times10^{-4}$ \\
\bottomrule
\end{tabular}
\caption{The resonant scale swept past a fixed mode by varying the energy,
$\sigma = 3\times10^{-3}$, $\lambda = 100$~km, domain carried to $5\times10^5$~km so that the L
resonance is crossed at every energy. The response turns on between $25$ and $30$~MeV, against
$E_c = 29.9$~MeV. The sub-threshold entries are at the $10^{-7}$ floor of this measurement; the
quoted error is the scatter over seeds about one noise-free run and excludes that run's own
integration error, so entries formally several standard errors from zero should be read as
floor.}
\label{tab:threshold}
\end{table}

\subsection{Amplitude}
\label{sec:amplitude}

Let $w_n(p,n_{\max}) = n^{-p}/\sum_{m \leq n_{\max}} m^{-p}$ be the fraction of the variance
carried by mode $n$ under the normalisation of~\eqref{eq:field}, and $r_n = R(n)/\sigma^2$ the
response per unit variance. If the modes act independently and quadratically in the amplitude,
\begin{equation}
  R(\sigma, p, n_{\max}) \;=\; \sigma^2
  \sum_{n \leq n_{\max}} w_n(p,n_{\max})\, r_n .
  \label{eq:compose}
\end{equation}
Both premises hold in this weak regime: the leakage scales as $\sigma^{2.02 \pm 0.03}$ over a
factor $33$ in amplitude, from $3\times10^{-4}$ to $10^{-2}$, with $R/\sigma^2$ drifting by $9\%$
from $40.4$ to $44.0$; and~\eqref{eq:compose}, built from the measured $r_n$, reproduces direct
broadband measurement within errors ($1.09 \pm 0.23$ for $p = 5/3$ with $16$ realisations,
$1.42 \pm 0.67$ for the two flanks with $8$). Since $r_n$ is appreciable only for $n$ of order a
few, moving the cutoff from $n_{\max}$ to $n_{\max}' > n_{\max}$ at fixed total amplitude
changes the response by
\begin{equation}
  \frac{R(n_{\max}') - R(n_{\max})}{R(n_{\max})} \;\simeq\; -f,
  \qquad f \;=\; \frac{\sum_{n = n_{\max}+1}^{n_{\max}'} n^{-p}}
                          {\sum_{n = n_{\min}}^{n_{\max}'} n^{-p}},
  \label{eq:law}
\end{equation}
minus the fraction of the variance reassigned to the added modes. The relation is exact if
those modes contribute nothing to the response, and approximate otherwise.

\begin{table}[h]
\centering
\begin{tabular}{clrrr}
\toprule
$p$ & regime & $\sigma$ & measured & predicted $-f$ \\
\midrule
$3$   & 2D enstrophy cascade       & $3\times10^{-3}$ & $+0.02 \pm 0.01\%$ & $-0.00\%$ \\
$5/3$ & Kolmogorov                 & $3\times10^{-3}$ & $-1.29 \pm 0.80\%$ & $-1.37\%$ \\
$1$   & shallower than any cascade & $3\times10^{-3}$ & $-21.5 \pm 12.5\%$ & $-21.47\%$ \\
\midrule
$5/3$ & Kolmogorov                 & $3\times10^{-4}$ & $-1.59 \pm 0.59\%$ & $-1.37\%$ \\
$1$   & shallower than any cascade & $3\times10^{-4}$ & $-15.5 \pm 12.3\%$ & $-21.47\%$ \\
\bottomrule
\end{tabular}
\caption{Equation~\eqref{eq:law} against measurement, cutoff moved from $n_{\max} = 200$ to
$1000$ at fixed total amplitude, $n_{\min} = 1$; for $p = 3$ the predicted change is
$-0.001\%$. The last two rows are at the amplitude
Borriello \emph{et al.}~\cite{borriello2013} measure, the Kolmogorov row there using $64$
realisations and the others $8$.}
\label{tab:law}
\end{table}

For $p = 5/3$ and $p = 1$ the measured changes agree with the prediction within their errors
(Table~\ref{tab:law}); in the fourth row the neglected contribution is
$(-6.9 \pm 2.3)\times10^{-8}$ on a total of $(4.33 \pm 0.72)\times10^{-6}$. For $p = 3$ the
predicted change, $-0.001\%$, is below the precision of the measurement, which gives
$+0.02 \pm 0.01\%$: small, but not a test at the $10^{-5}$ level. Equation~\eqref{eq:law} holds at fixed total amplitude. With the
coefficients of the resolved modes preserved and the continuation appended, as Borriello
\emph{et al.}~\cite{borriello2013} do, the same move changes the response by
$+0.02 \pm 0.13\%$ against $-1.36 \pm 0.23\%$ at fixed amplitude (twenty-four realisations each,
same phases): the modes between $0.5$ and $0.1$~km contribute no response here. Our cutoff sits
a factor $130$ below $\Losc$; the randomisation of the reference begins at scales of
$\mathcal{O}(100)$~km, above $\Losc$, so this test does not reach that prescription.

\subsection{Turbulence amplitude at the resonance}
\label{sec:turbulencia}

The response falls as $\lambda^{1.4}$, so what matters is the variance below $\Losc$, and that
variance is not in the data: at the resonance radius the grid of the archive's models has a
spacing of $155$ to $190$~km, so the shortest wavelength any of them contains is $4.7$ to
$5.7\,\Losc(20~\mathrm{MeV})$. Writing the variance per decade as $V(\lambda) = \tfrac12 P(n)\,n$
and fitting $V \propto \lambda^{q}$ over the half decade between the resampling roll-off and the
detrending filter, stacked over all rays with a single resonance crossing, gives $q = 2.29$,
with a range of $2.14$ to $2.48$ across four progenitors. Extrapolated below $\Losc$ with that
slope, the contrast is $\sigma = 7.2\times10^{-4}$ in the median, from $4.9\times10^{-4}$ to
$1.7\times10^{-3}$ between progenitors, and folding it through the modal response gives a leakage
of $8.8\times10^{-5}$, from $4.2\times10^{-5}$ to $4.5\times10^{-4}$; relative to $\Pee$, a median
of $3.9\times10^{-3}$. The amplitude is not stable between progenitors and is an extrapolation
below the grid. Whether the resonance lies inside the post-shock turbulent region depends on
which front is used: by the largest density drop it lies ahead of the front along $66\%$ of the
rays of M15-7b, by the velocity discontinuity behind it along $72\%$, with median separations of
$+159$ and $-153$~km, one grid zone either way. Measured on both sides, the unshocked material
carries more contrast in three of four progenitors.

At the amplitude of the synthetic field, $64$ realisations give
$R = (4.33 \pm 0.72)\times10^{-6}$, $1.9\times10^{-4}$ of $\Pee$; at $\sigma = 10^{-2}$, the ``few
per cent'' assumed by earlier idealised treatments, the response is $20\%$ of the signal. Fogli
\emph{et al.}~\cite{fogli2006} showed that fluctuations behind the front at an assumed amplitude
of a few per cent can partly erase the shock imprint; Borriello
\emph{et al.}~\cite{borriello2013} measured the amplitude in a high-resolution two-dimensional
model and found no significant damping; the quadratic response explains why a factor $33$ in
assumed amplitude is a factor $10^3$ in effect. The most recent treatment of turbulence and
flavour conversion \citep{bao2026} adopts a fluctuation amplitude of $0.4$ as ``a canonical
local normalization'' and asks that, in simulation-calibrated applications, it be ``extracted
along individual trajectories'', in the gain region rather than at the H resonance. In the
framework of Kneller and collaborators \citep{kneller2010,kneller2013,kneller2015spectrum}, the
H channel is relatively immune below a contrast of about $1\%$, effects grow rapidly above it,
and spectral information reaches the detector in the weak limit; an exponent of $2$ turns the
factor $14$ in amplitude between $1\%$ and $7\times10^{-4}$ into a factor $190$ in effect.
Kneller \& Volpe~\cite{kneller2010} also report sensitivity to amplitudes near $10^{-5}$ from
the combination of turbulence with multiple H resonances, which the single-crossing profile
here excludes by construction.

\subsection{A statistical caution}
\label{sec:estat}

The leakage is strongly non-self-averaging: over independent turbulence realisations its
scatter is $134\%$ of its mean, so absolute effects converge only as $N^{-1/2}$ from a wide
distribution, and eight realisations determine an absolute effect to about $50\%$. Differences
between cutoffs run at the same phases are paired and cancel most of that scatter, which is why
a $7\times10^{-8}$ contribution is measurable at $3\sigma$ with $64$ realisations while the
$4\times10^{-6}$ effect containing it is measured to $17\%$. An unpaired version of the
comparison of Table~\ref{tab:law} produced an apparent $40\%$ discrepancy in
Eq.~\eqref{eq:compose} that vanished with better statistics.

\section{The propagator, and its validation}
\label{ap:validacao}

The propagator is {\sc GHOST} \citep{ghost}. Its public repository contains the numerical core
that builds the Hamiltonian of Eq.~\eqref{eq:H} and evolves it, the command-line program, the
{\sc SNEWPY} interface, and acceptance tests that check the Hamiltonian, the matter term and
the conversion against closed-form results: two-flavour oscillation in vacuum, the MSW resonance at
constant density, three flavours with a CP phase, and the adiabatic crossing of a $\rho \propto
r^{-3}$ envelope.

\subsection{Mass convention in the matter term}
\label{sec:massa}

The matter term of Eq.~\eqref{eq:H} uses the neutron mass, which is the convention of the
profiles generated for this work. Writing it with the atomic mass unit at fixed $\rho$ and $Y_e$
would \emph{raise} the potential by $0.87\%$, since $m_u < m_n$, and lower every resonance
density and band edge by the same fraction. A profile from another source should be converted
so that it reproduces the intended electron density, rather than by the name of the mass unit.

\subsection{Requirements on the domain}
\label{sec:dominio_ap}

The domain must end below the L resonance for the highest energy used --- terminating where
$\rho = 11.7$~g\,cm$^{-3}$ leaves the L resonance outside the domain at $80$~MeV and produces an
error growing to $26\%$ --- and start well above the H resonance for the lowest. A convergence
scan on the parametrised profile at $t = 1$~s and $20$~MeV gives $\Pee = 0.05293$ for
$\rho(r_0)/\rho_H > 950$, $0.05308$ at a ratio of $118$, and $0.09310$ at a ratio of $35$, which
taken at face value asks for $\rho(r_0)/\rho_H > 10^3$. The experiments here ran at: a ratio of
$1600$ to $6300$ for the {\sc SNEWPY} validation profile; $403$ for the closed-form comparison
below; $50$ at $20$~MeV and $13$ at $5$~MeV for the perturbative scans of
Appendix~\ref{ap:perturbativo}; and $10^2$ after extension for the three-dimensional profiles,
whose files offer $113$ to $1.2$. On the three-dimensional profiles the result flattens once the
ratio exceeds $10$ and does not move out to $3\times10^3$, so the $76\%$ change at a ratio of
$35$ above is a property of the inner edge of the parametrised profile.

At the outer end, the perturbative scans were first run to $10^5$~km, where
$\rho = 10$~g\,cm$^{-3}$, while the L resonance sits near $20$~g\,cm$^{-3}$ at $20$~MeV and
scales as $1/E$, so at $40$ and $80$~MeV it falls at or beyond the end of that domain. Repeating
the measurement with the domain carried to $5\times10^5$~km, where $\rho = 0.08$~g\,cm$^{-3}$,
and pairing by seed, the response falls by $6\%$ at $40$~MeV ($8.8\sigma$) and by $20\%$ at
$80$~MeV ($5.0\sigma$): the truncated domain inflated it. Table~\ref{tab:threshold} is quoted on
the extended domain. At $20$~MeV, on fields that respond --- the single modes $n = 2$ and $4$ and
the broadband field at $p = 5/3$, $n_{\max} = 200$ --- the shifts are $-1.4\%$ ($3.4\sigma$),
$-1.6\%$ ($3.5\sigma$) and $-2.2\%$ ($1.8\sigma$), all the same way, so the perturbative results
carry a domain systematic of one to two per cent, inside the $9\%$ drift of $R/\sigma^2$ across
the amplitude scan; we record it rather than absorb it.

\subsection{Comparison with a closed form}
\label{sec:validacao}

For an exponential density profile the crossing probability is known in closed form
\citep{petcov1988,kuo1989}. With
$\gamma_{\rm res} = (\dmsq \sin^2 2\theta_{13}/2E\cos 2\theta_{13})\,H$ and
$H = |d\ln n_e/dr|^{-1}_{\rm res}$, the linear Landau--Zener form is
$P_H = e^{-\pi\gamma_{\rm res}/2}$, while for an exponential profile
\begin{equation}
  P_H \;=\;
  \frac{e^{-\frac{\pi}{2}\gamma_{\rm res}F} - e^{-\frac{\pi}{2}\gamma_{\rm res}F/\sin^2\theta_{13}}}
       {1 - e^{-\frac{\pi}{2}\gamma_{\rm res}F/\sin^2\theta_{13}}},
  \qquad F = 1 - \tan^2\theta_{13}.
\end{equation}
On profiles spanning $e^{\pm 6}$ in density about the resonance, reading $P_H = 1 - m_3$, the
solver reproduces the exponential result with a median deviation of $1.8\times10^{-3}$ over $P_H$
from $0.970$ to $0.010$, against $1.3\times10^{-2}$ for the linear form.

\subsection{A consistency test against solar data}
\label{sec:solar}

Borexino reports the electron-neutrino survival probability at four energies spanning a factor
of thirty \citep{borexino2021}, across the transition from the vacuum-averaged regime to the
adiabatic one. Taking the electron density from the standard exponential approximation,
$n_e/N_A = 245\,e^{-10.54\,r/R_\odot}$~cm$^{-3}$ \citep{bahcall1988}, starting each reaction at
its peak production radius and reporting $\Pee$, with the mixing parameters of
Section~\ref{sec:evolucao} and nothing fitted:

\begin{table}[h]
\centering
\begin{tabular}{lrrrr}
\toprule
reaction & $E$ (MeV) & $r_{\rm prod}/R_\odot$ & this work & Borexino \\
\midrule
pp         & $0.267$ & $0.10$ & $0.545$ & $0.57 \pm 0.09$ \\
$^7$Be     & $0.862$ & $0.07$ & $0.517$ & $0.53 \pm 0.05$ \\
pep        & $1.44$  & $0.06$ & $0.482$ & $0.43 \pm 0.11$ \\
$^8$B      & $8.1$   & $0.05$ & $0.308$ & $0.37 \pm 0.08$ \\
\bottomrule
\end{tabular}
\caption{The solar survival probability with an approximate density profile and no free
parameters, against the measured values. The deviations are $-0.28$, $-0.27$, $+0.47$ and
$-0.78\sigma$.}
\label{tab:solar}
\end{table}

The pp value approaches the vacuum-averaged limit
$\cos^4\theta_{13}\,(1 - \tfrac{1}{2}\sin^2 2\theta_{12}) + \sin^4\theta_{13} = 0.553$, and the
$^8$B value sits near the adiabatic $\sin^2\theta_{12}\cos^2\theta_{13} = 0.297$. Rescaling the
normalisation of the profile by $\pm30\%$ moves the four computed values by $0.03$, $0.24$,
$0.21$ and $0.24$ of the respective experimental errors. This is a consistency test of the chain
from mixing parameters through the matter potential to a survival probability, with an
approximate solar model and measurement errors much larger than the benchmarks above; it
exercises the adiabatic L resonance and says nothing about the sudden regime.

\section{Energy scan in M15-7b at 3 s}
\label{ap:energia}

\begin{table}[h]
\centering
\begin{tabular}{rrrrr}
\toprule
$E$ (MeV) & $w/\Losc$ & $r_{\rm res}/r_{\rm front}$ & raw & reconstructed \\
\midrule
 5 & $23.6$ & $1.01$ & $1.018$ & $6.25$ \\
10 & $11.8$ & $1.01$ & $1.028$ & $\bm{10.38}$ \\
20 & $5.9$  & $1.02$ & $1.029$ & $6.87$ \\
40 & $2.9$  & $1.26$ & $1.027$ & $2.30$ \\
60 & $2.0$  & $1.44$ & $1.002$ & $1.18$ \\
\bottomrule
\end{tabular}
\caption{Twenty rays of M15-7b per energy, $t = 3$~s, a separate run from that of
Table~\ref{tab:orcamento}. The last two columns are $\Pee$ in units of the adiabatic
prescription. $w/\Losc$ is the numerical front width at each energy, located by the largest
density drop, as the reconstruction step locates it; by the velocity discontinuity of
Table~\ref{tab:m157b} each entry is $0.71$ times smaller. $r_{\rm res}/r_{\rm front}$ locates
the resonance relative to the shock.}
\label{tab:energia}
\end{table}

\bibliographystyle{JHEP}
\bibliography{refs}

\end{document}

%% file: ph_tabela.tex
\begin{tabular}{rrrrl}
\toprule
$t$ (s) & $E$ (MeV) & $P_H$ raw & $P_H$ reconstructed & 10--90\% over rays \\
\midrule
$3$ & $5$ & $0.000$ & $\bm{0.544}$ & $0.05$--$0.74$ \\
$3$ & $10$ & $0.000$ & $\bm{0.808}$ & $0.74$--$0.84$ \\
$3$ & $20$ & $0.001$ & $\bm{0.482}$ & $0.08$--$0.85$ \\
$3$ & $40$ & $0.000$ & $\bm{0.019}$ & $0.00$--$0.30$ \\
$3$ & $60$ & $0.000$ & $\bm{0.004}$ & $0.00$--$0.04$ \\
\midrule
$10$ & $5$ & $0.019$ & $\bm{0.019}$ & $0.00$--$0.12$ \\
$10$ & $10$ & $0.002$ & $\bm{0.002}$ & $0.00$--$0.01$ \\
$10$ & $20$ & $0.000$ & $\bm{0.001}$ & $0.00$--$0.00$ \\
$10$ & $40$ & $0.000$ & $\bm{0.000}$ & $0.00$--$0.00$ \\
$10$ & $60$ & $0.000$ & $\bm{0.001}$ & $0.00$--$0.02$ \\
\bottomrule
\end{tabular}